\documentclass[12pt]{article} 

\usepackage{hyperref}
\usepackage{url}
\usepackage{setspace}
\usepackage{amsmath,amssymb,amscd,mathrsfs}
\usepackage{amsfonts}
\usepackage{color}
\usepackage{graphicx}
\usepackage{braket}
\usepackage[english]{babel}
\usepackage{bbold}
\usepackage{cite}

\usepackage[applemac]{inputenc}
 \usepackage[toc,page]{appendix}

\usepackage{verbatimbox}
\usepackage{fancyvrb}
\usepackage{tensor}
\usepackage{array,multirow,makecell}
\usepackage{eqnarray}
\usepackage{float}
\usepackage{animate}

 \def\p{\partial}
 \newcommand{\bea}{\begin{eqnarray}}
\newcommand{\eea}{\end{eqnarray}}
\newcommand{\be}{\begin{equation}}
\newcommand{\ee}{\end{equation}}
\newcommand{\ba}{\begin{align}}
\newcommand{\ea}{\end{align}}

\newcommand{\EE}{\mathcal{E}}
\newcommand{\xp}{x^+}

\newcommand{\xpsq}{x_\perp^2}
\newcommand{\xperp}{\vec{x}_\perp}
\newcommand{\sP}{s^+}
\newcommand{\sm}{s^-}
\newcommand{\spsq}{s_\perp^2}
\newcommand{\sperp}{\vec{s}_\perp}
\newcommand{\vp}{v^+}
\newcommand{\vm}{v^-}
\newcommand{\vpsq}{v_\perp^2}
\newcommand{\vperp}{\vec{v}_\perp}

\newcommand{\pp}{\partial_+}
\newcommand{\bp}{\Box_\perp}

\newcommand{\eps}{\epsilon}
\newcommand{\difop}[1]{\left(\sum_{q_{#1},r_{#1},s_{#1},t_{#1}} c_{q_{#1},r_{#1},s_{#1},t_{#1}}^{#1}(x_{12}\cdot \xi_{-})^{q_{#1}}(\partial\cdot \xi_{+})^{r_{#1}} (\partial \cdot \xi_{-})^{s_{#1}}(\partial\cdot\partial)^{t_{#1}}\right)}
\newcommand{\cond}[1]{-q_{#1}+r_{#1} + s_{#1} + 2 t_{#1}}
\newcommand{\conp}[1]{q_{#1} - r_{#1} + s_{#1}}

\numberwithin{equation}{section}

\def \e {\epsilon}

\newcommand\rref[1]{(\ref{#1})}
\newcommand{\AB}[1]{{\color{red}{{\bf AB:} #1}}}

\newlength{\slength}
\begin{document}

\begin{titlepage}
\begin{center}

\hfill \\
\hfill \\
\vskip 0.75in

{\Large \bf  Einstein gravity from  ANEC correlators}\\

\vskip 0.4in

{\large Alexandre Belin$^{a,b}$, Diego M. Hofman$^a$, Gr\'egoire Mathys$^a$ } \\
\vskip 0.3in

${}^{a}${\it Institute for Theoretical Physics, University of Amsterdam,
Science Park 904, Postbus 94485, 1090 GL Amsterdam, The Netherlands} \vskip 0.5mm
${}^{b}${\it CERN, Theory Division, 1 Esplanade des Particules, Geneva 23, CH-1211, Switzerland}
\vskip 3.5mm

\texttt{a.belin@cern.ch, d.m.hofman@uva.nl, g.o.mathys@uva.nl}

\end{center}

\vskip 0.75in

\begin{center} {\bf ABSTRACT } \end{center}
We study correlation functions with multiple averaged null energy (ANEC) operators in conformal field theories. For large $N$ CFTs with a large gap to higher spin operators, we show that the OPE between a local operator and the ANEC can be recast as a particularly simple differential operator acting on the local operator. This operator is simple enough that we can resum it and obtain the finite distance OPE. Under the large $N$ - large gap assumptions, the vanishing of the commutator of ANEC operators tightly constrains the OPE coefficients of the theory. An important example of this phenomenon is the conclusion that $a=c$ in $d=4$. This implies that the bulk dual of such a CFT is semi-classical Einstein-gravity with minimally coupled matter.

\vfill

\noindent \today

\end{titlepage}

\tableofcontents

\pagebreak

\section{Introduction}

The AdS/CFT correspondence \cite{Maldacena:1997re} relates conformal field theories (CFTs) in $d$ dimensions to gravitational theories in Anti de Sitter spacetimes in $d+1$ dimensions (AdS$_{d+1}$).  This duality was originally discovered in the context of string theory by considering D-brane constructions, which relate particular CFTs that arise in the low energy limit of brane theories to string theory in AdS. When the gravitational dual is described by weakly coupled Einstein gravity, the CFTs are pushed into some extreme corners of theory space, with a large number of degrees of freedom and a strong coupling. 

A more modern approach to the subject is to consider the strongest form of the AdS/CFT correspondence which states that every conformal field theory can be viewed as giving rise to a theory of quantum gravity in AdS. Typically, however, the bulk dual will be highly curved and quantum. One then tries to answer the following questions: what makes a CFT holographic? What are necessary and sufficient conditions that a CFT must satisfy in order to have a weakly coupled Einstein gravity dual? How does the bulk emerge from the CFT? These questions have mostly been tackled following two main frameworks. The first is the conformal bootstrap, originally conceived in \cite{Ferrara:1973yt,Polyakov:1974gs,Belavin:1984vu} and revived and modernized in \cite{Rattazzi:2008pe}. Here the idea is to use the analytic properties of CFT correlation functions to determine the physics of the dual gravitational theory. The second is that of quantum information theory where one might try  to derive Einstein's equations from CFT entanglement, see e.g. \cite{Ryu:2006bv,Lashkari:2013koa,Faulkner:2013ica,Faulkner:2017tkh,Lewkowycz:2018sgn}. The current work discussed in this note will follow the first approach to the problem, even though the full power of the bootstrap will not be exploited. 

The implementation of the conformal bootstrap in this context was originally introduced by Heemskerk, Penedones, Polchinski and Sully \cite{Heemskerk:2009pn}. They showed that there was an equivalence between the number of solutions to the crossing equations and bulk effective Lagrangians in AdS. Furthermore, they conjectured that the locality of the bulk theory was encoded in the dimension of the lowest single-trace operator with spin greater than two. Since then, much evidence has been gathered in favour of this conjecture, as well as a more quantitative understanding of the effect of the gap, see e.g. \cite{ElShowk:2011ag,Afkhami-Jeddi:2016ntf,Kulaxizi:2017ixa,Meltzer:2017rtf,Costa:2017twz,Camanho:2014apa}. There has also been a great amount of progress by bootstrapping partition functions in the large $N$ limit \cite{Keller:2011xi,Keller:2014xba,Hartman:2014oaa,Benjamin:2015vkc,Belin:2016yll,Belin:2016knb,Shaghoulian:2016xbx,Belin:2017nze,Mefford:2017oxy,Belin:2018oza,Hellerman:2009bu,Friedan:2013cba,Qualls:2014oea,Collier:2016cls,Keller:2017iql,Cho:2017fzo,Anous:2018hjh,Afkhami-Jeddi:2019zci}.

The statement that the bulk theory should be weakly coupled is really a statement about the three-point function of the CFT stress-tensor, suitably normalized. 

\begin{equation}
      \frac{\left\langle T T T\right\rangle}{\braket{TT}^{3/2}}=\vcenter{\hbox{\includegraphics[width=0.3\linewidth]{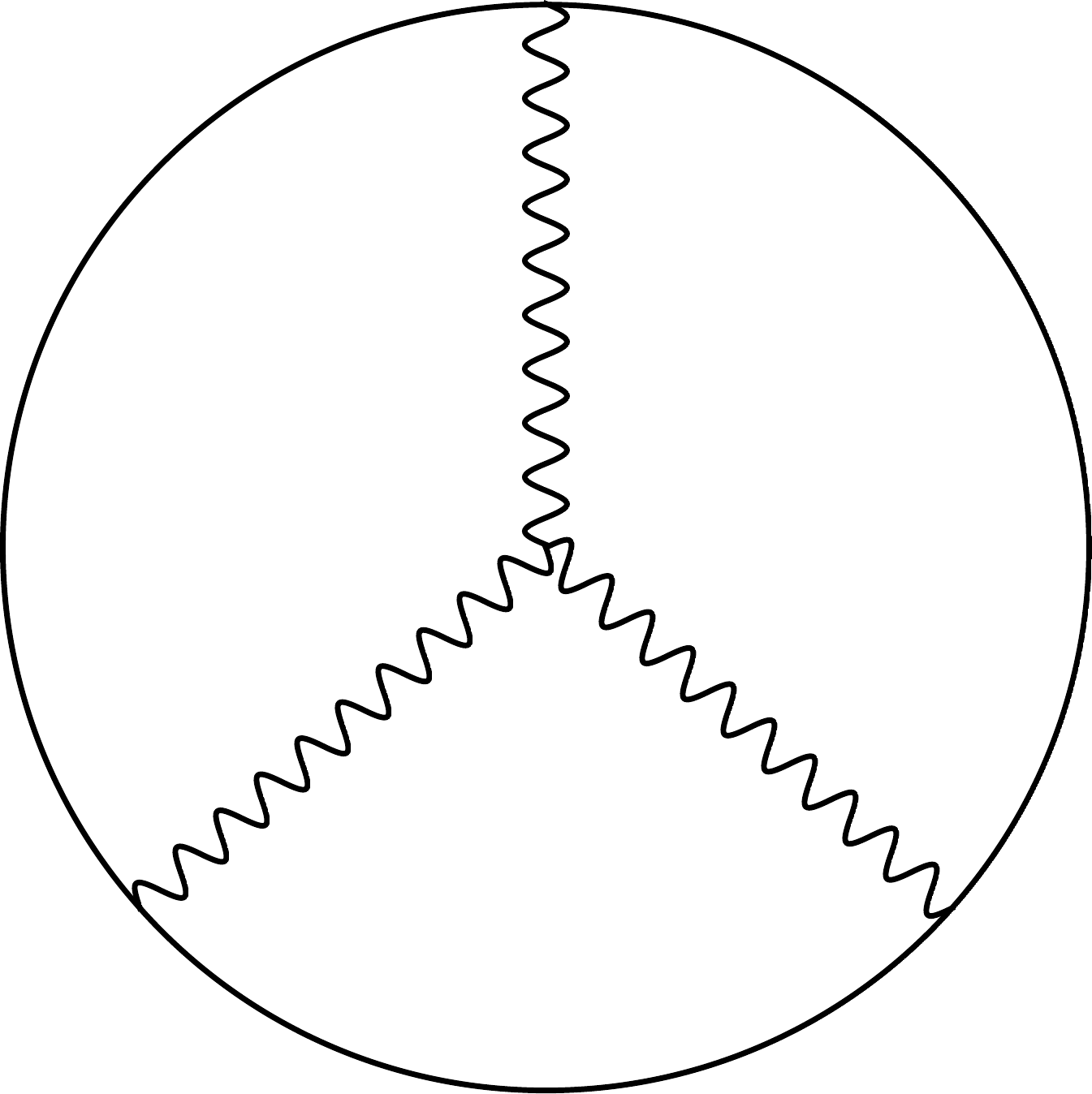}}} \sim \sqrt{G_N}
    \end{equation}

In two dimensions, there is a single tensor structure for the stress-tensor three-point function, and the normalized three-point function scales as $c^{-1/2}$, where $c$ is the central charge. 2d CFTs with a weakly-coupled gravity dual should therefore have a large central charge. In higher dimensions, there are multiple tensor structures for the stress-tensor three-point function. Concretely, in $ d \geq 4$ dimensions, we have three independent structures:
\be \label{TTT}
\braket{TTT}= t_0 \left[\braket{TTT}_0 +t_2 \braket{TTT}_2  + t_4 \braket{TTT}_4 \right]
\ee
where the $\braket{TTT}_i$ are known tensor structures given in \cite{Osborn:1993cr}. The index $i$ labels the irreducible representations appearing in the three-point function in terms of its spin. The $t_i$'s are theory-dependent structure constants and one would expect them all to be large for the gravitational theory to be weakly coupled. 

On the other hand, a generic bulk effective Lagrangian will be
\be\label{eftgrav}
S_{\text{bulk}}= \frac{1}{16\pi G_N} \int \sqrt{|g|} d^{d+1}x \left(-2\Lambda + R + \frac{\alpha_1}{\Lambda} R^2 +  \frac{\alpha_2}{\Lambda^2} R^3\right) +\cdots
\ee
\noindent where $\Lambda$ is dimensionful and $\alpha_1$ and $\alpha_2$ are dimensionless constants. The $\ldots$ represent terms suppressed by powers of $G_N$. The statement made above about the largeness of the $t_i$'s translates to the fact that we expect a hierarchy of scales that  separates $\Lambda$  from $G_N$ while keeping $\alpha_1$ and $\alpha_2$ order one. The reason we could isolate just these terms in the effective action is that we can use field redefinitions to codify the information of the three-point functions discussed above in just these terms.

An immediate concern that arises from these expressions is that they seem to violate the logic of effective field theory where we expect all higher derivative terms to be suppressed by the cutoff scale $G_N$ and not the IR scale $\Lambda$. This amounts to a large amount of fine tuning. But since holography allows this as a consequence of the large $N$ limit there is no obvious contradiction.

Eventually, causality considerations in the gravitational theory showed that only a finite, order one, range for the dimensionless constants $\alpha_1$ and $\alpha_2$ was allowed. This was first observed in \cite{Brigante:2007nu,Brigante:2008gz} and complete bounds were obtained in \cite{Buchel:2009tt,Hofman:2009ug}. It turns out these bounds follow directly from an exact calculation in CFT at finite $N$ performed in \cite{Hofman:2008ar} where bounds on the central charges were obtained by assuming a form of the averaged null energy condition in the context of a gedanken collider experiment. We review these results briefly in the following subsection.

Still, after these results were understood, it remained surprising that the effective field theory results could be violated by the possibility of finite values for $\alpha_1$ and $\alpha_2$. This puzzle was beautifully resolved in \cite{Camanho:2014apa} where a careful study of micro-causality in high energy scattering processes in the bulk of AdS showed that in a theory with a large gap $\alpha_1$ and $\alpha_2$ must effectively vanish. In this work, an argument was presented that deviations from this result are constrained by the dimension ($\Delta_{gap}$) of the lightest operator with spin $J>2$ in the dual CFT. In particular,

\be
\left| \alpha_1 \right| \lesssim \frac{1}{\Delta_{gap}^2}.
\ee
For example, one cannot have a CFT that is dual to Gauss-Bonnet gravity with a large correction to the Einstein-Hilbert action where all other higher derivative corrections are small. Through \rref{TTT}, this implies the vanishing of $t_2$ and $t_4$. It is known that these coefficients are related to values of central charges. The result is that holographic CFTs (i.e. large N, large gap) must satisfy in $d=4$:

\be
a=c \,,
\ee
where $a$ and $c$ are the anomaly coefficients. This result was later proven using CFT techniques in \cite{Afkhami-Jeddi:2016ntf}. The argument is technically quite involved, it requires an understanding of the Regge limit and rests ultimately on the chaos bound \cite{Maldacena:2015waa}.

The goal of this paper is to derive this result by simpler arguments in a large N CFT by assuming that there is a large gap to the higher spin operators. We will make heavy use of the averaged null energy operator, discussed in the next subsection, and its commutation relations, see \cite{Casini:2017roe,Cordova:2018ygx}. We hope this approach will open simpler ways to access the question of possible deviations from $a=c$ in terms of $\Delta_{gap}$ and sheds light on the sub-algebra of light-ray operators in general and their universal properties.

\subsubsection*{The averaged null energy operator}

The averaged null energy operator (which we will call the ANEC operator) is defined to be
\be
\EE(x^+,\xperp) = \int dx^- T_{--}(x^+,x^-,\xperp) \, ,
\ee
where $x^{\pm}$ are null directions. It is an example of a larger class of non-local CFT operators known as light-ray operators, see \cite{Balitsky:1987bk,Braun:2003rp} and  \cite{Kravchuk:2018htv,Cordova:2018ygx} for more recent developments. This operator has remarkable properties, in particular its expectation value is positive for any state in the Hilbert space
\be
\bra{\psi} \EE \ket{\psi} \geq 0 \,.
\ee
This inequality is known as the averaged null energy condition (ANEC) and is an astonishing property of quantum field theory. It has recently been proven in \cite{Faulkner:2016mzt,Hartman:2016lgu} by different methods. Interestingly, the two proofs originate again from either quantum information theory or the bootstrap. Positivity of the ANEC operator was assumed to derive bounds on the anomaly coefficients in any 4d conformal field theory in \cite{Hofman:2008ar}.  One simply evaluates the one-point function of an ANEC operator in a state created by the stress-tensor. The bounds read
\be\label{bounds}
 \frac{31}{18} \geq \frac{a}{c} \geq \frac{1}{3} \,.
\ee
These bounds are usually referred to as the \textit{conformal collider bounds} as they were first suggested in \cite{Hofman:2008ar} in the context of the gedanken collider experiment. They were first proven in \cite{Hofman:2016awc} by Lorentzian bootstrap techniques \cite{Hartman:2015lfa,Hartman:2016dxc}. The saturation of these bounds implies that the CFT is actually free, see \cite{Zhiboedov:2013opa}. 

Now that the ANEC has been proven to be true the bounds \rref{bounds} follow trivially as a particular instance of a much more general statement. One could generate many more inequalities using these type of techniques. For example, bounds on the OPE coefficients of other operators, e.g $\braket{TTO}$, were obtained in \cite{Cordova:2017zej}.

Once defined, the ANEC operator can be used to compute higher point correlation functions of the type
\be
\bra{\psi} \EE_1... \EE_k \ket{\psi} \,,
\ee
\noindent which is a $k+2$-point correlation function in the CFT. Notice that, if inserted at the same light-like coordinate $x^+$ but at different transverse positions $\xperp$, these operators are space-like separated and, therefore, should commute \cite{Casini:2017roe,Cordova:2018ygx}. However, this argument is in fact too quick since the two ANEC operators still touch at infinity, as can be seen on the Penrose diagram of Minkowski space. One could even perform a conformal transformation to map the point where they touch to a finite distance, making the concern more manifest. Nevertheless, while the null integral of generic operators may be problematic, one can show that the commutator still vanishes for ANEC operators (see \cite{Kologlu:2019bco} for an in-depth discussion of the issue).

Quite trivially, the product of commuting positive operators is also positive. This signals that once the ANEC is satisfied no further information in terms of bounds should be accessible from the higher point correlation functions.

These operators were related in \cite{Hofman:2008ar} by a conformal transformation to light-ray operators inserted at the conformal boundary of Minkowski space. In this context their interpretation corresponds to the insertion of a detector that measures the integrated energy deposited in a calorimeter in the celestial sphere in a sort of collider experiment. Because of the properties of the ANEC operators, these energy operators are also known to be positive and commuting.\footnote{In this context, the concern about commutation arises from a possible contribution coming from the point of future infinity in the Penrose diagram.}  

The energy operators can be obtained by sending the ANEC operator to the boundary of space-time as:

\be \label{defe}
E(n^i) \sim \lim_{r\to\infty}r^2 \int dx^-_{n} \, T_{--}(x^+_n= 2 r, x^-_n, 0)\, ,
\ee
\noindent where $n^i$ is a unit vector in space transforming under $SO(3)$. We have picked here a coordinate system $x^\pm_n = t \pm n^i x^i$ in order to take the limit in a simple way. Given this connection we will be a bit careless in the rest of this work and refer to these operators as energy or ANEC operators indistinctively.   

Higher point correlators of these operators can also be studied. In this work, we will study such correlation functions for generic CFTs at large $N$, and pay particular attention to the situation where there is a large gap for the single-trace higher spin operators. To do so, we will develop the OPE between an ANEC operator and a local operator. When there is a large gap, the only single-trace operator that will appear in the OPE between $\EE$ and $O$ is the local operator $O$ itself. In such a case, the ANEC operator acts as a differential operator which takes a relatively simple form.

We will show that the OPE expansion can be resummed to obtain an exact expression at finite distance between the ANEC operator and the local operator insertion. This will allow us to compute the conformal collider higher point correlation functions for a large $N$ CFT. We will see that consistency of the commutator of energy operators $E$ or, equivalently, demanding that their product remains positive singles out the dual AdS bulk theory to be Einstein Gravity.

From the point of view of the large $N$ CFT what we observe is that the range of allowed central charges \rref{bounds} is drastically reduced. By looking at commutators or higher point functions of energy correlators, we deduce
\be\label{sbounds}
1 \geq \frac{a}{c} \geq 1.
\ee
Concretely we show:

\begin{alignat}{6}
&\textbf{Result 1:}& \qquad &\bra{T} [E_1,E_2] \ket{T} &&= 0 \qquad  &&\Longrightarrow \qquad &a&=c \notag \\
\phantom{00} \notag \\
&\textbf{Result 2:}& \qquad   &\bra{T} E_1... E_n \ket{T} &&\geq0 \qquad  &&\Longrightarrow  \qquad &|a-c|&\leq \delta_n \,,
\end{alignat}
where $\delta_n$ are strictly decreasing. We will also derive equivalent properties for the coupling to currents. From a bulk point of view, we are deriving minimal Einstein-gravity couplings by assuming the large gap condition.

The result \rref{sbounds} was obtained, as mentioned above, first in \cite{Afkhami-Jeddi:2016ntf} by different methods. We expect our discussion to be somewhat simpler and also allow for applications of this techniques to other problems. A direct application, which we do not exploit in this work, is the possibility, within our formalism, to compute arbitrary higher point correlators of ANEC operators at finite distances in large $N$ CFTs. 

Ultimately, what allows these computations is that, as a consequence of the large $N$ limit and the large gap, higher dimensional CFTs acquire, as far as the ANEC operators go, a structure reminiscent of that found in 2d CFTs, where the Virasoro algebra fixes all correlation functions. Interesting examples where similar structures have been uncovered in higher dimensional CFTs include:  infinite dimensional algebras in supersymmetric theories \cite{Beem:2013sza,Beem:2014kka}, BMS algebras for light-ray operators \cite{Cordova:2018ygx} and affine Kac-Moody algebras for theories with higher form symmetries \cite{Hofman:2018lfz}. We hope our formalism will provide a tool to further study these results.

The paper is organized as follows: in section 2, we start by performing a warm up calculation in 2d CFTs, from which we draw analogies to higher dimensions. In section 3, we compute the OPE between local operators and the ANEC operator and use it to obtain an expression for the energy operator in terms of a differential operator. In section 4, we compute higher-point functions of the ANEC operators and derive constraints on the central charges of the CFT. We show the holographic dual theory must be Einstein gravity minimally coupled to other light fields. We conclude in section 5 with a discussion.

\textbf{Note added:} while this paper was in preparation, we learned about  \cite{Kologlu:2019bco} which contains some overlap with the results discussed here.

\section{A 2d warmup}

We start by considering the ANEC operator in two dimensions. As we will see this example is somewhat trivial, but it will nevertheless illustrate some important properties that will carry over to higher dimensions. In two dimensions, the ANEC operator is simply
\be
\EE=\int dz \,T(z) \,,
\ee
where $T(z)$ is the holomorphic stress-tensor. Let us start by considering the three-point function
\be
\braket{T(z_1)\EE T(z_3)} \,.
\ee
The local three-point function is given by
\be
\braket{T(z_1)T(z_2)T(z_3)}=\frac{c}{z_{12}^2z_{13}^2z_{23}^2} \,,
\ee
where $z_{ij} = z_i - z_j$. We can easily integrate this expression to obtain\footnote{We use an $i\eps$ prescription such that we pick up the poles that lie in the lower half plane and we close the contour in that direction, see more details about this in section 3.}
\be
\braket{T(z_1)\EE T(z_3)}=2\pi i \frac{2c}{z_{13}^5} \,.
\ee
However, there is a quicker way to arrive at this answer. In two dimensions, the ANEC operator is simply a differential operator
\be\label{diff2}
\EE=\int dz \,T(z) \sim \p_z \,,
\ee
which follows from the Laurent expansion of the stress-tensor. Thus, we have
\be
\braket{T(z_1)\EE T(z_3)}=-2\pi i \p_{z_1} \braket{T(z_1) T(z_3)} =2\pi i \frac{2c}{z_{13}^5} \,,
\ee
in agreement with the previous calculation. Having understood this, one can easily compute a $k$-point function of ANEC operators, by applying the differential operator $k$ times
\be
\braket{T(z_1)\EE_1..\EE_k T(z_2)} \sim \p^k \braket{T(z_1) T(z_2)} \,,
\ee
which means, in particular, that it is fixed by symmetry. While this is natural in light of Virasoro symmetry, it is quite surprising from a global conformal group point of view, where for example a four-point function is given by an infinite sum over the exchange of quasi-primaries. Let us illustrate this fact by considering the four-point function
\be
\braket{T(z_1) \EE_2 \EE_3 T(z_4)} \,.
\ee
This correlator, much like the local four-point function, is completely fixed by symmetry in two dimensions due to the Ward identities \cite{Belavin:1984vu}. Let us start from the canonically normalized stress-tensor two-point function
\be
\langle T(z)T(w) \rangle = \frac{c/2}{(z-w)^4}\label{eq:TT2Pt} \, .
\ee
By using the Ward identity (see appendix \ref{subsec:TPhi}) twice, we get the local four-point function
\bea \label{4ptTWI}
\langle T(z_1)T(z_2)T(z_3)T(z_4)\rangle &=&
\frac{c^2/4}{z_{12}^4z_{34}^4}+
   \frac{c^2/4}{z_{23}^4
 z_{14}^4}+\frac{c^2/4}{z_{13}^4
 z_{24}^4}\\
   &+&\frac{2 c}{z_{12}^2 z_{23}^2
z_{14}^2z_{34}^2}-\frac{2 c}{z_{12}z_{13}^2
z_{23}z_{14}z_{24}^2z_{34}}\nonumber \,. 
\eea
We can now integrate this four-point function twice. The terms on the first line of \rref{4ptTWI}, which are crucially proportional to $c^2$, do not have simple poles and therefore vanish. The answer comes solely from the terms on the second line and reads
   \begin{align}
\braket{T(z_1) \EE_2 \EE_3 T(z_4)}=\int \int \langle T(z_1)T(z_2)T(z_3)T(z_4)\rangle dz_2 dz_3=& -(4\pi^2)\frac{10c}{(z_1-z_4)^6}.\label{eq:Fin2DWI}
   \end{align}
We would now like to rederive this result from the point of view of a global conformal block expansion, where one sums over an infinite set of quasi-primaries.
  \subsubsection*{Using conformal blocks}
 We want to explicitely check the computation of the double integral above using the conformal block decomposition. Note that it corresponds to the exchange of an infinite number of quasi-primaries. These operators are the stress tensor itself, as well as composites of the stress-tensor of the schematic form $:T \p^k T:$ which are also quasi-primaries. We will now show that the only global block relevant for computing the integrated four-point function is the stress-tensor one. All the composites will drop out once we integrate. We will use the results of \cite{Osborn:2012vt}. The four-point function can be written as 
 \begin{align}
 \langle T(z_1)T(z_2)T(z_3)T(z_4)\rangle = \frac{1}{z_{12}^4z_{34}^4}\mathcal{F}_{TTTT}(\eta),
 \end{align}
where the cross-ratio is defined as 
   \be 
   \eta = \frac{z_{12}z_{34}}{z_{13}z_{24}}.
   \ee
The result \eqref{4ptTWI} can be repackaged in terms of the cross-ratio as  
 \be 
 \mathcal{F}_{TTTT}(\eta) = \frac{1}{4}c^2\left(1+\eta^4+\frac{\eta^4}{(1-\eta)^4}\right)+2c\eta^2\frac{1-\eta+\eta^2}{(1-\eta)^2},
 \ee 
 which can be written as a sum of global conformal blocks as 
 \be 
  \mathcal{F}_{TTTT}(\eta) = \frac{1}{4}c^2 + \sum_{p=0}^{\infty}a_{2p}\eta^{2p+2}F(2p+2,2p+2;4p+4;\eta),\label{TTTTCB}
 \ee
 where $F(a,b;c;z)$ is the hypergeometric function and the coefficients $a_{2p}$ are
 \be 
 a_{2p}= \left(\frac{1}{144}c^2(2p-1)_6 + 2c(1+2p(2p+3))\right)\frac{(2p)!(2p+1)!}{(4p+1)!}.
 \ee
For our purposes, we can drop the constant piece coming from the identity operator. It is the first term in \rref{4ptTWI} and we saw it vanishes when we integrate. We can thus consider
\begin{align}  \label{Ftilde}
 \tilde{\mathcal{F}}_{TTTT}(\eta) = \mathcal{F}_{TTTT}(\eta)- \frac{1}{4}c^2 =&  \sum_{p=0}^{\infty}a_{2p}\eta^{2p+2}F(2p+2,2p+2;4p+4;\eta).
 \end{align}
Let us look at the expansion of $\tilde{\mathcal{F}}_{TTTT}(\eta)$ for small $\eta$. Taking into account the factor of $\frac{1}{z_{12}^4z_{34}^4}$, the series will contain terms, for a given $m$, of the form 
 \be 
 \frac{1}{z_{12}^4z_{34}^4}\eta^m = \frac{1}{(z_{12}z_{34})^{4-m}}\frac{1}{(z_{13}z_{24})^m}.
 \ee
It is clear that when taking the integral as $z_1\rightarrow z_2$, only terms where $0\leq  m\leq 3$ can contribute since otherwise there is no pole at $z_1=z_2$. By inspecting \rref{Ftilde}, it is clear that the only block for which this happens is $p=0$ which is the exchange of the stress-tensor. The $p=0$ block gives a contribution
\begin{align}
\int\int\left.\frac{1}{{z_{12}^4z_{34}^4}} \tilde{\mathcal{F}}_{TTTT}(\eta)\right|_{p=0} dz_2dz_3= -(4\pi^2)\frac{10c}{(z_1-z_4)^6},
\end{align}
in agreement with the full answer. All other blocks corresponding to the composites vanish once we integrate. This is quite remarkable: an infinite number of blocks are needed to reproduce the local four-point function but a single one survives the integrals. In a large $c$ theory, the composites $:T \p^kT:$ can be thought of as double-trace operators. Even though Virasoro symmetry does not prevail in higher dimensions, we will see that the double-trace operators also drop out of correlators with the ANEC. This is the first lesson to draw from this simple two dimensional case. Secondly, the fact that the ANEC operator is a differential operator in $d=2$ \rref{diff2} is no longer true in higher dimensions. However, we will see that in a large $N$ theory with a large gap, it becomes approximately true again. The differential operator is slightly more complicated, but it involves a finite number of minus derivatives. With this newly gained insight, we are ready to discuss the structure of ANEC operators in higher dimensions.

%There are two important lessons to draw from this example. First, the ANEC operator has become a differential operator, which truncates to a finite number of $z$-derivatives (in this case a single derivative). A more complicated version of this statement will be true in higher dimensions.
%
%Second, there were terms in the local four-point function of stress-tensors that scaled like $c^2$. From a global conformal block point of view, these correspond to the exchange of the operators $:T^2:$.\footnote{We give the details of the 2d calculation in appendix \ref{2d}.} The contribution of such operators has completely dropped out. At large $c$, such operators should be thought of us as double-trace operators, and they do not contribute to the integrated correlation function. While in two dimensions this statement is exact in $c$, it will still be true at large $N$ in higher dimensions. This will be of key important for the rest of the paper. We will now derive an OPE between the ANEC operator and local operators in higher dimensions. 

\section{Action of $\EE$ on local operators \label{opelocal}}
We now study correlation functions of the ANEC operator in higher dimensions. While our technology should apply to any dimension $d>2$, we will work in four dimensions for the rest of the paper. The goal of this section is to develop the OPE between the ANEC operator $\EE$ and any local operator $O_{\mu_1...\mu_s}$ of spin $s$. In the spirit of the previous section, we will recast the ANEC operator as a differential operator. This will be an exact statement at the level of CFT three-point functions, which are fixed by symmetry. While most results in this section are well known, the advantage of this formalism is that under certain assumptions on the CFT (large $N$, large gap), the differential operator we find will enable us to compute higher-point correlation functions with multiple ANEC operators, which will be the subject of the following section. For now, we restrict to three-point functions. We will start by explaining the general structure of this differential operator and then we move on to concrete calculations for operators of different spin.

\subsection{ANEC operator as a differential operator \label{6rules}}

The goal of this section is to show that the ANEC operator can be recast as a differential operator. At the level of three-point functions with two identical local operators, this expression is exact. We will obtain
\be
\braket{O\EE O} = \mathcal{D} \braket{OO} \,,
\ee
for some differential operator $\mathcal{D}$. In the above expression $O$ is a local operator of arbitrary spin. This differential operator follows directly from considering the expansion of the OPE between $\EE$ and $O$ as we now discuss. Throughout this section, we only care about the terms in the OPE where $O$ itself appears as that is the only relevant information for this three point function. In the next section we will argue why in large $N$ theories with a gap this is all we need even for higher point-functions.

We start by discussing the building blocks of the differential operator. Consider a spacelike vector $n^i$ of unit norm. We define two null coordinates as
\be
x^{\pm}=t\pm n^i x^i \,,
\ee
and define two associated null vectors $\xi_{\pm}$. The two spacelike coordinates will be denoted $\xperp$. This gives a natural decomposition of the Lorentz symmetry as
\be
SO(1,3) \to SO(1,1)\times SO(2) \,.
\ee
\noindent where the $SO(2)$ leaves $n^i$ invariant. This splitting corresponds to separating the two null directions and the two space-like directions. To proceed, pick a null vector ( say $\xi^\mu_-$) that will point in the direction along which we will integrate the stress tensor component $\xi^\mu_- \xi^\nu_- T_{\mu \nu}$.  Then, we need the spin information of the local operator. If we have an operator of spin $s$, we will consider its contraction with a polarization tensor
\be
\mathbb{O} = \epsilon^{\mu_1...\mu_s} O_{\mu_1...\mu_s} \,.
\ee
We can now build the most general differential operator that respects the symmetries at hand. The rules for the OPE $\mathcal{E}(x_1) \mathbb{O}(x_2) $ are the following:
\begin{enumerate}
\item  It is a scalar, so all indices must be contracted.
\item It is built from the constituents: $\xi_+^\mu, \ \xi_-^\mu,  \ x_{12}^\mu= x^\mu_1 - x^\mu_2, \ \epsilon^{\mu_1 ... \mu_s}, \ O^{\mu_1 ... \mu_s}, \ g^{\mu\nu}, \ \p^\mu$.
\item It is linear both in the operator and in the polarization tensor.
\item The position vector $x_{12}^\mu$ can only be contracted with $\xi_-$\footnote{We are going to look at the OPE when the operators are separated only in $+$ direction. The more general result can be obtained by  $SO(1,3)$ transformations.}.
\item The differential operator must have weight 3.
\item It must carry a $SO(1,1)$  $-$  index.
\item If the operator is a conserved current, the derivative operator cannot be contracted with $O^{\mu_1 ... \mu_s}$. If it is traceless, the operator cannot be contracted with the metric either.
\end{enumerate}
For example, consider the case of a scalar operator. The most general operator that we can write down under the conditions above is
\be
\EE O = \sum_{q,r,s,t}c_{q,r,s,t} \left(x_{12}  \cdot \xi_-\right)^q \left(\p \cdot \xi_+\right)^r \left(\p \cdot \xi_-\right)^s  \left(\p \cdot \p\right)^t O \,,
\ee
with
\be
-q+r+s+2t = 3 \,, \qquad q-r+s=1 \,,
\ee
and where $c_{q,r,s,t}$ are some coefficients that are unfixed for now. The constraints on $q,r,s,t$ follow directly from the rules above. Furthermore, this expansion must be local at short distances. This means that $r, s, t$ are positive integer powers. We can then rewrite the differential operator as\footnote{In later sections, it will sometimes be more convenient to use a slightly different basis: we will use $\bp$ instead of $\p^\mu \p_\mu$ which is simply a reshuffling of the basis elements written here.}
\bea \label{generalopscalar}
\mathcal{D}= \frac{1}{\left(x_{12}  \cdot \xi_-\right)}\sum_{k=0}^{\infty} &\Big(& a_k \left(\p \cdot \xi_-\right)^2 +b_k \left(x_{12}  \cdot \xi_-\right)\left(\p \cdot \xi_-\right) \left(\p \cdot \p\right)   \notag \\
&+&c_k \left(x_{12}  \cdot \xi_-\right)^2\left(\p \cdot \p\right)^2   \Big) \left( x_{12}  \cdot \xi_- \ \p \cdot \xi_+\right)^k  \,.
\eea
The coefficients $a_k,b_k,c_k$ can be computed explicitly by comparing with the three-point function. We will do this in detail in the next subsection. Similar expressions exist for operators with spin and we write down the explicit expression for $U(1)$ currents and the stress tensor, see  Appendix \ref{diffopspin}. 

Before deriving the precise differential operator \rref{generalopscalar} for scalars, let us first discuss the properties of the differential operator once we send it infinitely far away to the celestial sphere, following \rref{defe}. This object will be relevant when we compute correlation function of energy detectors in scattering experiments. By symmetry considerations, we are able to restrict the form of the diferential operator, up to a few coefficients that may be extracted once \rref{generalopscalar} is known. This is just a rephrasing of the well known analysis of  \cite{Hofman:2008ar}.

\subsubsection{The large distance limit and matrix elements}

\begin{figure}[h!]
\centering
\includegraphics[width=0.55\textwidth]{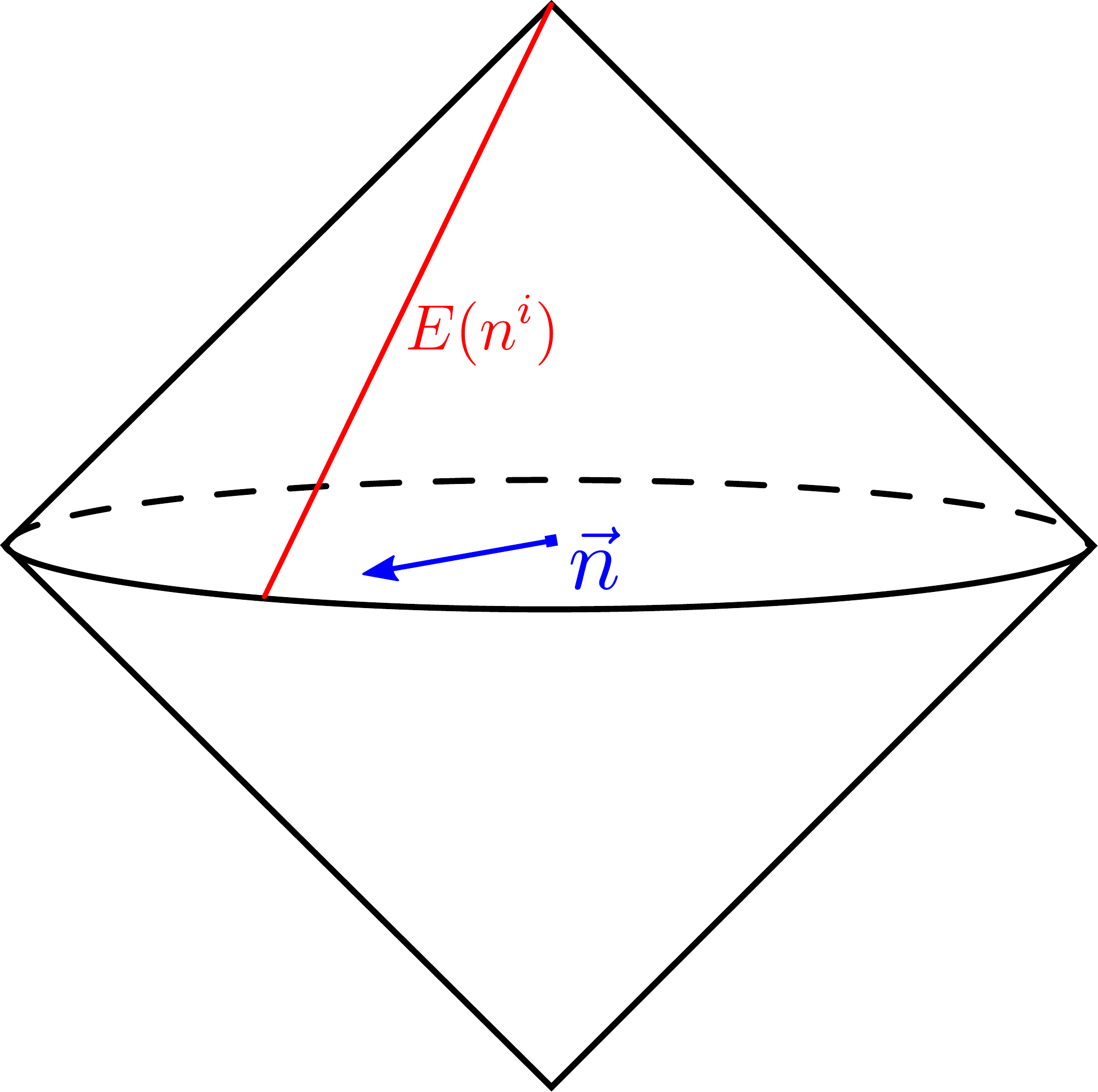}
\caption{The Penrose diagram of Minkowski space in three dimensions. The ANEC operator has been sent to spatial infinity, at a given point on the celestial sphere (here a point on the circle). It is still integrated along a null direction represented by the red line. The vector $\vec{n}$ indicates the direction in which the operator is inserted.}
\label{cone}
\end{figure}

We would like to view these correlation functions as scattering experiments with the ANEC operators being energy detectors \cite{Hofman:2008ar}. To do so, it is important to send the ANEC operators to spatial infinity, inserted at a given angle on the celestial sphere. This is illustrated in Fig. \ref{cone}. From this point of view, it is therefore more useful to split the Lorentz symmetry as
\be
SO(1,3) \to \mathbb{R} \times SO(3) \,.
\ee
After the limit, and an appropriate rescaling by $r^2$, our ANEC operator has undergone the limit \rref{defe} and we now refer to it as an energy operator $E(n^i)$.
We are now interested in the matrix elements
\be \label{genmatrixel}
\frac{\bra{\mathbb{O}} E(n^i) \ket{\mathbb{O}}}{\braket{\mathbb{O}|\mathbb{O}}} \,.
\ee
It is useful to go to momentum space so we consider the Fourier transform of the two-point function
\be \label{twopF}
F(q) \equiv \int e^{i q \cdot x_{12}}\braket{\mathbb{O}(x_2)|\mathbb{O}(x_1)}  \,,
\ee
and we will be particularly interested in the four-momentum
\be  \label{0k}
q=(q^0,0,0,0) \,,
\ee
namely $\vec{q}=0$ momentum eigenstates. For such states, the matrix elements become extremely simple. By dimensional analysis, the differential operator will necessarily be proportional to the energy $q^0$. We will give explicit expressions in the following subsection. For now, consider the general form of the energy density one-point function \cite{Hofman:2008ar}. We have
\be 
\frac{\bra{\mathbb{O}} E(n^i) \ket{\mathbb{O}}}{\braket{\mathbb{O}|\mathbb{O}}} =  \frac{q^0}{4\pi \epsilon^*\epsilon} \epsilon^*M(g^{ij},n^i, C_{OOT})\epsilon \,,
\ee
where $C_{OOT}$ represents the relevant OPE coefficients and our notation is
\bea
\eps^*\eps&\equiv&\eps_{i_1...i_s} \eps^{i_1...i_s} \notag\,, \\
\eps^*M\eps&\equiv& \eps_{i_1...i_s} M^{i_1...i_s}_{\phantom{i_1...i_s} j_1...j_s} \eps^{j_1...j_s}\,.
\eea
From this point of view, one should really view the ANEC operator as a transfer matrix $M$ between the polarizations. The only important detail is that this must be considered in an in-in formalism. As we made manifest above, it depends on the OPE coefficients between the stress tensor and the operator creating the state\footnote{For operators with spin, note that there is more than one OPE coefficient.} and its indices are built from the unit vector and the metric. For scalar operators, we have no polarization and the energy is uniformly spread over the sphere, namely $M=1$. For operators with spin, the situation is more interesting. For example, the transfer matrices for $U(1)$ currents and stress-tensors read \cite{Hofman:2008ar}
\bea \label{transferMatrix}
\eps^* M_J \eps&=&  \eps^*_i \left[g^{ij} + a_2 \left(n^i n^j-\frac{1}{3}g^{ij}\right)\right] \eps_j \, ,\\
\eps^* M_T \eps&=&\eps^*_{ij} \left[ g^{ik} g^{jl} + t_2 g^{ik}\left( n^j n^l -\frac{ g^{jl}}{3} \right) + t_4 \left(n^in^jn^kn^l- \frac{2g^{ik} g^{jl}}{15}  \right) \right] \eps_{kl}\,.
\eea
where the constants $a_2,t_2,t_4$ depend on the OPE coefficients and the coefficient of the two-point function. We work them out in detail below by comparing with the OPE. For operators of spin $s$, the transfer matrix carries $2s$ indices and its explicit dependence on the OPE coefficients can be worked out in a similar fashion. 

These expressions, including the values of the coefficients $a_2,t_2,t_4$ can be obtained explicitly by resumming the expression for the differential operator of the type  \rref{generalopscalar} which has an infinite radius of convergence as we show below. 

At this point, it is worthwhile to pause and ask why we should go through the trouble of computing this differential operator, since we could have obtained \rref{transferMatrix} directly from the three-point functions. It turns out, that because of the infinite radius of convergence of the OPE expansion within this subsector, we can use these expressions to compute higher order correlation functions of the form
\be \label{genmatrixel}
\frac{\bra{\mathbb{O}} E(n_1^i)...E(n_k^i) \ket{\mathbb{O}}}{\braket{\mathbb{O}|\mathbb{O}}} \,.
\ee
These could even be generalized to finite distance in flat space, away from the celestial sphere where the symmetry considerations are no longer directly applicable.

We will see in the following section that under the large gap assumption, this higher point matrix element is completely determined by the transfer matrix we have just described, once we take the appropriate limit. To see this, note that when the ANEC can be recast as a differential operator in higher point functions, then the product of ANECs is simply the product of the differential operators. Upon taking the appropriate limit, equation \rref{genmatrixel} becomes
\be
\frac{\bra{\mathbb{O}} E(n_1^i)... E(n_k^i) \ket{\mathbb{O}}}{\braket{\mathbb{O}|\mathbb{O}}} = \left( \frac{q^0}{4\pi}\right)^k  \frac{1}{\epsilon^* \cdot \epsilon}\epsilon^* M_1...M_k \epsilon \,.
\ee
This will provide an efficient tool to obtain bounds on the OPE coefficients. We will now discuss the details of extracting the differential operator for scalar operators as well as the way to take the large distance limit.

\subsection{Scalar operators}
In this section, we derive the differential operator when it acts on scalar local operators, namely  \rref{generalopscalar}. The derivation for operators with spin follows but is more tedious. We give some of the details of the computation for a current and the stress-tensor in Appendix \ref{diffopspin}. We will see that we can give the form of the operator not only as a series expansion, but also in a compact resummed expression. From this, it becomes very simple to extract the large distance limit and derive the transfer matrix described above, as we  explain below.

\subsubsection{Exact result}
We start with the general expression for the three-point function of the stress-tensor with two scalar operators \cite{Osborn:1993cr}
\be \label{TOO}
\braket{T_{\mu\nu}(x_1)O(x_2)O(x_3)}= \frac{C_{TOO}}{x_{12}^dx_{13}^dx_{23}^{2\Delta-d} }t_{\mu\nu}(X) \,,
\ee
where in $4$ dimensions
\be
C_{TOO}= - \frac{2\Delta}{3\pi^2}  \,,
\ee
and
\be 
t_{\mu\nu}(X) = \frac{X_{\mu}X_{\nu}}{X^2}-\frac{1}{d}\delta_{\mu\nu},\qquad X_{\mu} = \frac{(x_{12})_{\mu}}{x_{12}^2}-\frac{(x_{13})_{\mu}}{x_{13}^2}.
\label{tmunu}\ee
It will be more convenient for us to define
\be
s^{\mu}=x_{12}^\mu \,, \qquad v^{\mu}=x_{23}^\mu \,,
\ee
which also gives $x_{13}^\mu=s^\mu+v^\mu$. The three-point function is only a function of these two vectors by translational invariance. We will always think of the vector $s^\mu$ as being the vector between the ANEC operator $\EE$ and $O$ and $v^\mu$ being the vector between the two scalars. We now wish to compute
\be
\braket{\EE(x_1) O(x_2)O(x_3)} \,.
\ee
To do this, we need to specify time orderings for the operators, which is done by giving the appropriate $i\e$ prescription. We thus perform the shifts
\be
x_i^{\pm} \to x_i^{\pm} + i \e_i.
\ee
We refer the reader to Appendix \ref{notation} for our conventions for the $\pm$ notation.

We will perform the $x^-$ integral by means of a contour integral in the complex $x^-$ plane. From that point of view, the role of the $i\e$ prescription will be to determine which poles lie inside our contour. To obtain a non-vanishing answer, we need to have solely the singularity due to $O(x_2)$ in our contour and not that of $O(x_3)$ (or the other way around). This can be achieved by picking 
\be
\e_{12}>0 \,, \qquad \e_{13}<0 \,,
\ee
and closing the contour through the bottom. This corresponds to a time ordering where we first create the state with $O(x_2)$, then insert $\EE$ and then we go back in time to $O(x_3)$ to create another in state.

We can now compute the correlator with the ANEC operator, obtained by integrating \rref{TOO}. We have
\be
\braket{O \EE O}= \int ds^- \frac{-\frac{2\Delta}{3\pi^2}}{s^d(s+v)^dv^{2\Delta-d} }t_{--}(X) \,.
\ee
The integrand has a pole at
\be
s^-= \frac{\spsq}{s^+}-i \e_{12} \frac{\spsq+(\sP)^2}{(\sP)^2}+\mathcal{O}(\e_{12}^2) \,,
\ee
and the integral can be computed using Cauchy's theorem by considering the residue at the pole. Upon further taking the limit $\e_{12},\e_{13}\to0$, we obtain
\be \label{TOOexact}
\braket{O(x_3) \EE(x_1) O(x_2)}=(-2\pi i)  \frac{\Delta}{\pi^2} \frac{(\vp)^2 (1+\frac{\sP}{\vp})^2}{\sP v^{2\Delta+4} (1- \frac{(\sP)^2\vm -2\sP\sperp \cdot\vperp+\vp \spsq}{\sP v^2})^3} \,.
\ee
We will now show how to reproduce this answer using an OPE expansion.
\subsubsection{The OPE expansion}
We will now consider the OPE expansion of the ANEC operator with a scalar operator. The OPE between a scalar operator and the stress tensor is given by
\be
T_{\mu\nu}(x_1) O(x_2) = A_{\mu\nu} O(x_2) +B_{\mu\nu}^{\alpha}\p_\alpha O(x_2) + C_{\mu\nu}^{\alpha\beta}\p_\alpha \p_\beta O(x_2) + ...
\ee
We give  the explicit expression for the tensors in appendix \ref{TOOPE}, but they are basically the most general tensors built out of the vector $s_{\mu}$ and the metric $g_{\mu\nu}$ that satisfy the basic symmetry properties of the indices, namely symmetric traceless in $\mu,\nu$ and symmetric in $\alpha,\beta,...$. The coefficient in front of every tensor structure can be extracted by matching to the expansion of the three-point function \rref{TOO}.

Note that only the tensors $A,B,C,..$ carry an $s^{\mu}$ dependence, so we can perform the $\sm$ integral term by term to extract the OPE between $\EE$ and $O$.\footnote{There are many simplifications that occur once we take the integral. For example, it is easy to see that only the tensor structures with at most two metric factors can contribute once we take the $s^-$ integral. Having three or more metric factors would be accompanied by enough powers of $s$ to cancel the pole at $s^2=0$.} We will now compare the expressions that we get at each order in the OPE expansion to the exact answer \rref{TOOexact}. It is easy to see that the contributions from $A_{\mu\nu}$ and $B_{\mu\nu}^{\alpha}$ drop out and the first contribution to the three-point function comes from $C_{\mu\nu}^{\alpha\beta}$ and yields
\be
(-2\pi i)  \frac{\Delta}{\pi^2} \frac{(\vp)^2}{\sP v^{2\Delta+4}} \,.
\ee
At the next order, the contribution from $D_{\mu\nu}^{\alpha\beta\gamma}$ is
\be
(-2\pi i)  \frac{\Delta}{\pi^2} \frac{v^{+}}{(s^{+})^2v^{2\Delta+6}} \left(-6 \sP \vp \sperp \cdot \vperp+ 3(\vp)^2\spsq+(\sP)^2 (2\vpsq+\vm\vp) \right) \,.
\ee
At this point, it is already easy to understand where this expression comes from by looking at the exact result \rref{TOOexact}. Rewriting the full answer as
\be 
\braket{O(x_3) \EE(x_1) O(x_2)}=(-2\pi i)  \frac{\Delta}{\pi^2} \frac{(\vp)^2 (1+\beta_1)^2}{\sP v^{2\Delta+4} (1-\beta_2)^3} \,.
\ee
with
\bea \label{xandy}
\beta_1&=&\frac{\sP}{\vp}\,, \notag \\
\beta_2&=&\frac{(\sP)^2\vm -2\sP\sperp \cdot\vperp+\vp \spsq}{\sP v^2} \,.
\eea
The k-th order in the OPE gives the homogeneous polynomial of order $k-2$ in the $\beta_{1,2}$ expansion of the exact answer. This shows that the ANEC operator can be recast as a differential operator acting on the scalar operator. This is an exact statement at the level of three-point functions. We will now proceed to write this operator explicitly. We will do so for  $\sperp=0$, which will be enough for our purposes. The full expression for $\sperp \neq 0$ can be recovered by using $SO(1,3)$ transformations to change coordinates. For states created by operators with spin, we will no longer be able to fix the polarization vectors since most of the rotational symmetry has been used to align the ANEC operator in the $\sP$ direction. This will render the expressions slightly more complicated but the concept remains the same.

\subsubsection{The explicit form of the differential operator}

It is straightforward to work out the differential operator by integrating the OPE between the scalar and the stress-tensor. Writing the operator in the form \rref{generalopscalar} we find\footnote{It is  more convenient to use $\bp$ rather than $\p^2$, which is a change of basis.}
\bea \label{Dseries}
\mathcal{D}&=&  (-2\pi i) \frac{\Delta}{\pi^2} \frac{1}{\sP} \sum_{k\geq0}\left(\frac{a_k}{\Delta_{k+2}}\p_-^2+\sP  \frac{b_k}{\Delta_{k+3}}\p_- \bp+(\sP)^2\frac{c_k}{\Delta_{k+4}}\bp^2\right) \left(\sP \pp\right)^k  \, , \quad\quad
\eea
with
\bea
a_k&=&1\,, \notag \\
b_k&=&\frac{k+1}{2}\,,\\
c_k&=& \frac{k^2+3k+2}{32} \,,
\eea
and where we have used the Pochhammer symbol
\be
\Delta_x = \frac{\Gamma(x+\Delta)}{\Gamma(\Delta)} \,.
\ee
Fortunately, we can explicitly resum the operator. We find
\bea \label{difoper}
\mathcal{D}&=&  (-2\pi i) \frac{\Delta}{\pi^2} \Bigg[  \frac{\p_-^2}{ \sP}\frac{e^{\sP \p_+}}{(\sP \p_+)^{\Delta+1}}  \frac{\Gamma(\Delta+1) -\Gamma(\Delta+1,\sP\p_+)}{\Delta}\notag \\
&+& \frac{\p_-\Box_\perp}{2\Delta_2} \left( 1 + \frac{e^{\sP \p_+}(\sP\p_+ -\Delta-1)(\Gamma(\Delta+2)-\Gamma(\Delta+2,\sP\p_+))}{(\sP\p_+)^{\Delta+2}}\right) \\
&+& \frac{\sP\Box_\perp^2}{32\Delta_5}\left(\frac{\Delta_5}{\Delta_3}(\sP\p_+-\Delta)+\frac{e^{\sP\p_+}((\sP \p_+)^2 - 2\frac{\Delta_2}{\Delta} \sP \p_+ +\frac{\Delta_3}{\Delta})(\Gamma(\Delta+5)-\frac{\Delta_5}{\Delta_3}\Gamma(\Delta+3,\sP \p_+))}{(\sP\p_+)^{\Delta+3}}\right)  \Bigg] \notag \,,
\eea 
which is a relatively simple operator. Here $\Gamma(s,x)$ is the incomplete Gamma function. At this point, it is worth comparing this answer to the one we found in two dimensions. First, we see that the differential operator involves only a finite number of minus derivatives, as advertised. In two dimensions, the operator truncated to a single minus derivative. Here it is more complicated but the number of derivatives remains bounded. For operators with spin, it is also bounded. Second, we notice the appearance of an exponentiation of the plus derivative. This is a new feature compared to two dimensions.

There is also another phenomenon happening. The OPE expansion had a finite radius of convergence, given essentially by $\beta_2=1$ in \rref{xandy}. Now, in terms of the differential operator, the series can be resumed with infinite radius of convergence, as is manifest by the exponential factor in \rref{difoper}. This fact is reminiscent of Borrel resummation and is of importance in taking the large distance limit, over which we now have control. We are now ready to send the ANEC operator to the celestial sphere and consider energy correlators. The differential operator will simplify even further.

\subsubsection{The large distance limit}

We are now completely set up to study energy correlators. The  kinematic setup we are interested in consists of states that are created by inserting local operators near the center of Minkowski space. Furthermore, we wish to send the ANEC operator(s) far away in the radial direction (in these coordinates, in the $\xp$ direction). The limit corresponds to taking $\sP\to\infty$ and the leading term in the exact three-point function \rref{TOOexact} becomes
\be 
\braket{O(x_3) \EE(x_1) O(x_2)}_{\sP\to\infty}\sim \, (2\pi i) \frac{\Delta}{\pi^2} \frac{1}{(\sP)^2v^{2\Delta}}\frac{v^2}{ (\vm)^3} \,.
\ee
%This formula can be understood quite intuitively: the factor of $(\sP)^{-2}=1/4r^2$ corresponds to the surface area of the celestial 2-sphere. The factor of $v^{-2\Delta}$ will disappear once we normalize by the norm of the state. Finally, if we went to Fourier space, we would find a uniform energy distribution on the sphere which is what the factor of $\frac{v^2}{(v^-)^3}$ represents. This is because the state is created by a scalar local operator, and it will drastically change when we consider operators with spin.
One can study the differential operator $\mathcal{D}$ given in \rref{difoper} in this limit. We find that the operator becomes particularly simple
\be
\mathcal{D} \sim (-2\pi i) \frac{\Delta}{\pi^2}  \frac{1}{(\sP)^2} \left( -\frac{1}{\Delta}\frac{\p_-^2}{\p_+}+\frac{1}{2\Delta} \frac{\bp \p_-}{\pp^2}-\frac{1}{16\Delta} \frac{\bp^2}{\pp^3}\right) \,.
\ee
It is worthwhile to mention that there are two asymptotic behaviours for the incomplete regularized Gamma function and one of them contains an exponential. In the regime of real momenta that we are interested in, this exponential is a pure phase and it does not dominate the long distance limit. This is particularly clear when the differential operator acts on momentum eigenstates. Asymptotically, for states satisfying \rref{0k} :
\be
\mathcal{D}  \sim \frac{2 i}{\pi(\sP)^2}\frac{\p_-^2}{\p_+} \left( 1-\Gamma(\Delta+1)\frac{e^{\sP \p_+}}{(\sP \p_+)^\Delta} \right) =\frac{q^0}{\pi (\sP)^2} \left(1-\Gamma(\Delta+1) \frac{e^{-i \sP q^0/2}}{(-i \sP q^0/2)^\Delta}\right) \,.
\ee
From this is obvious that the second term is much smaller than the piece we have kept for any $\Delta >0$.

Now consider the Fourier transform of the two-point function \rref{twopF} where we use \rref{0k}
\be
F(q)= \int e^{i q \cdot x} \frac{1}{v^{2\Delta}} \, .
\ee
The action of the operator then becomes extremely simple and we find
\be
\mathcal{D}F(q) \sim (2\pi i) \frac{1}{\pi^2}  \frac{1}{(\sP)^2} \frac{\p_-^2}{\p_+}F(q)= \frac{q^0}{4\pi r^2}F(q) \,.
\ee
The energy operator \rref{defe} is extremely simple when acting on the momentum space two-point function. We then have (in momentum space)
\be\label{OE}
E(n^i) O = \frac{q^0}{4\pi} O \,.
\ee
Note that this is not an approximate formula, it is exact (provided we do not care about other operators appearing in the OPE). At the level of expectation values, we find
\be \label{scalar3p}
\braket{E(n^i)}\equiv \frac{\braket{O E O}}{\braket{OO}}=\frac{q^0}{4\pi} \,,
\ee
namely a uniform energy distribution on the celestial sphere as expected for scalar states. This will drastically change once we consider states built out of operators with spin, which we now discuss.
\subsection{Operators with spin}
For operators with spin, one repeats the same procedure in a straightforward fashion. The most general expression for the differential operator of the form \rref{generalopscalar} is given explicitly for conserved currents and the stress-tensor in \rref{diffopgencur} and \rref{diffopgenT}, respectively. One then compares the general expression with the direct computation of the integrated three-point function and extracts the values of the expansion coefficients. Note that there is some gauge freedom in the OPE because of the conservation/tracelessness of the operators but it can be dealt with reasonably painlessly. In practice, we pick a gauge that makes the resummation easier, which we can do without loss of generality.

Having the differential operator in the form \rref{Dseries}, we can simply perform the sum and obtain the resummed version which is a (much) lengthier version of \rref{difoper}. We omit the exact expression from this draft for environmental reasons. Once again, taking the large distance limit and acting on momentum eigenstates drastically simplifies the operators and we obtain the equivalent of \rref{OE} for operators with spin. Namely, 

\bea
E \epsilon^\mu J_\mu &=& \frac{\pi q^0}{4}\left(\frac{3(\tilde{c}-2\tilde{e})}{2c_v}\epsilon \cdot J -3\frac{\tilde{c}-8\tilde{e}}{c_v}\left(\xi_+ \cdot J\right)( \xi_+-\xi_-) \cdot \epsilon \right) \,, \\ 
E \eps^{\mu\nu} T_{\mu\nu} &=& \frac{\pi q^0}{4} \Bigg( \frac{5}{3}\frac{7\hat{a}+2\hat{b}-\hat{c}}{c_T} \eps^{\mu\nu}T_{\mu\nu} +10 \frac{13\hat{a}+4\hat{b}-3\hat{c}}{c_T} \xi_+^\mu T_{\mu\nu} \epsilon^{\nu\rho}(\xi^-_\rho-\xi^+_\rho) \notag \\
 &-&\frac{15}{6} \frac{81\hat{a}+32\hat{b}-20\hat{c}}{c_T} \xi_+^\mu\xi_+^\nu T_{\mu\nu} \epsilon^{\rho\sigma}(\xi^-_\rho-\xi^+_\rho)(\xi^-_\sigma-\xi^+_\sigma) \Bigg) \,.
\eea
From this, we can extract the transfer matrices \rref{transferMatrix} and we find well known expressions for the coefficients 
\bea \label{OPEcouplings}
a_2 &=&\frac{3(8\tilde{e}-\tilde{c})}{2(\tilde{c}+\tilde{e})}\,, \\
t_2 &=&\frac{30(13 \hat{a} +4\hat{b} -3\hat{c})}{14\hat{a}-2\hat{b}-5\hat{c}}\,, \\
t_4 &=&-\frac{15(81\hat{a}+32\hat{b}-20\hat{c})}{2(14\hat{a}-2\hat{b}-5\hat{c})} \,.
\eea
The tilde and hat coefficients correspond to the OPE coefficients appearing respectively in (3.13) and (3.19) of \cite{Osborn:1993cr}. One can relate the OPE coefficients to the anomaly coefficients $a$ and $c$ given by
\be
T^\mu_\mu= \frac{c}{16\pi^2} W^2-\frac{a}{16\pi^2}E \,,
\ee
where $W$ is the Weyl tensor and $E$ is the Euler density. The relation to the OPE coefficients is
\be \label{aoverc}
\frac{a}{c} = \frac{9\hat{a}-2\hat{b}-10\hat{c}}{3(14\hat{a}-2\hat{b}-5\hat{c})} \,.
\ee
The values of the OPE coefficients \rref{OPEcouplings} naturally agree with the results in \cite{Hofman:2008ar}, obtained directly from the integrated three-point function without going through the OPE.

 As explained in \cite{Hofman:2008ar}, the positivity of the ANEC operator for arbitrary polarizations yields the conformal collider bounds
\bea
-\frac{3}{2}&\leq& a_2  \ \ \leq 3 \,,\notag \\
0&\leq& 1-\frac{t_2}{3}-\frac{2t_4}{15} \,,\notag \\
0&\leq&t_2+ 2\left(1-\frac{t_2}{3}-\frac{2t_4}{15}\right)\,,\notag \\
0&\leq& t_2+t_4+ \frac{3}{2}\left(1-\frac{t_2}{3}-\frac{2t_4}{15}\right).
\eea
The last three-inequalities can be recast to constrain the anomaly coefficients $a$ and $c$ as
\be\label{acb}
 \frac{1}{3} \leq \frac{a}{c} \leq \frac{31}{18} \, .
\ee 
Now, the point is that in a consistent finite $N$ CFT this is the end of the story. No other information can be obtained by looking at higher point functions, as the energy operators at different positions commute and, therefore, their product is automatically positive.

It turns out that large $N$/large gap CFTs are somewhat sick, unless stronger constraints than \rref{acb} are imposed. We will see below that either by studying the commutator of energy operators (effectively a four point function in the CFT) or by looking at higher point functions we will obtain that the bounds above need to be strengthened to

\be
a_2 = t_2 = t_4 = 0 \,,\qquad  \Longrightarrow \quad \quad \frac{a}{c}=1\,.
\ee
Therefore, we now turn our attention to higher-point functions.

\section{Correlation functions of the ANEC operator}

We have seen that we can rewrite the ANEC operator as a differential operator, which is an exact statement at the level of three-point functions within the subspace of operators involved. We are now interested in computing correlation functions with multiple ANEC operators. For four-point functions and higher, this is a complicated task since all operators can run in the exchange channel and the four-point function therefore knows about the entire spectrum of the theory. We will focus on large $N$ theories, where large-$N$ factorization will give us a lot of mileage. We start by reviewing the properties of correlators at large $N$. Our counting will be adapted to theories that have order $N^2$ degrees of freedom like $\mathcal{N}=4$ SYM or adjoint theories in general. It is straightforward to adapt it to other types of large $N$ theories if needed.

The upshot of this section is that, for holographic CFTs, the computations from the previous section are enough to compute all higher point functions of ANEC operators. This, in turn, results in strong constraints for the OPE coefficients in the theory.

\subsection{Review of large $N$ factorization \label{largeN}}

In large $N$ theories, it is useful to separate the operators into light and heavy operators. Light operators have a conformal dimension $\Delta$ that is fixed as $N\to\infty$. From a gravitational point of view, these operators correspond to fields from the bulk effective field theory living in AdS. In particular, this excludes all operators that create black holes, as their conformal dimensions scales with $N$. We further separate these operators into two classes: single-trace and multi-trace operators. Single-trace operators correspond to bulk fields following the usual AdS/CFT dictionary whereas multi-trace operators correspond to multi-particle states of the bulk fields.

What distinguishes single-trace and multi-trace operators is the way correlation functions scale. This is easiest to see in the normalization where the single-trace operator is normalized such that
\be
\braket{OO} \sim \mathcal{O}(1) \,.
\ee
It is important to note that this is \textit{not} the canonical normalization for the stress-tensor or conserved currents, which typically have a two-point function that scales like $N^2$. We will review the case of the stress-tensor separately. The higher point functions are then given by
\be \label{largeNfac}
\braket{OOO}\sim \mathcal{O}\left( N^{-1}\right) \,, \qquad \braket{OOOO}_c \sim \mathcal{O}(N^{-2})\, ,
\ee
where $\braket{...}_c$ is the connected correlation function. On the other hand, the multi-trace operators satisfy
\be
\braket{ :OO: \ :OO:}\sim  \mathcal{O}(1) \,, \qquad \braket{ :OO: \ :OO: \  :OO:}\sim  \mathcal{O}(1) \,,
\ee
namely their correlation functions are order one (provided there exist Wick contractions) \cite{Belin:2017nze}.

For any two single-trace operators $O_1$ and $O_2$, there exists at large $N$ a family of double trace operators
\be
[O_1O_2]_{n,l} \sim O_1 \p_{\mu_1}...\p_{\mu_l} (\p^\nu \p_\nu)^n O_2 \,,
\ee
with conformal dimension (to leading order in $N$)
\be \label{DTdelta}
\Delta_{n,l}^{(0)}=\Delta_1+\Delta_2+2n+l \,.
\ee
Typically, the multi-trace operators give contributions at the largest order in the $1/N$ expansion. For example, consider the correlation function
\bea
\braket{O(x_1)O(x_2)O(x_3)O(x_4)} &=&\braket{O(x_1)O(x_2)}\braket{O(x_3)O(x_4)}+\braket{O(x_1)O(x_3)}\braket{O(x_2)O(x_4)} \notag \\ 
&+&\braket{O(x_1)O(x_4)}\braket{O(x_2)O(x_3)} +\mathcal{O}(1/N^2)\, . 
\eea
Now consider the conformal block expansion of the correlation function above, in the $1\leftrightarrow2$ and $3\leftrightarrow4$ channel. The exchange of the identity operator gives the first term, whereas all the double-trace operators sum up to give the other two Wick contractions \cite{Heemskerk:2009pn}. We will denote this contribution DT$^{(0)}$. Note that these double-trace operators only have the conformal dimensions \rref{DTdelta} at infinite $N$. Their dimensions (and OPE coefficients) get modified once we include $1/N^2$ contributions to the four-point function, as is required by crossing symmetry. In general, the structure is 
\bea \label{DT1}
\Delta_{n,l} &=& \Delta_{n,l}^{(0)} + \frac{1}{N^2}\gamma_{n,l}+\dots \notag \\
C_{OO[OO]_{n,l}}&=&a^{(0)}_{n,l} +\frac{1}{N^2}a^{(1)}_{n,l}+\dots
\eea
The leading order OPE $a^{(0)}_{n,l}$ coefficients are given in \cite{Heemskerk:2009pn}. 

In this paper, we will be interested in computing the connected four-point function, which is of order $1/N^2$. There are two types of contributions at this order:
\begin{itemize}
\item The exchange of all single-trace operators. We will denote this contribution schematically by ST.
\item The contribution coming from the anomalous dimensions and the correction to the OPE coefficients of the double-trace operators. We will denote these contributions schematically as DT$^{(1)}$.
\end{itemize}
The correction of the double-trace data has two separate origins. Part of it corresponds to quartic couplings in the bulk, and this part can be added freely in a crossing symmetric way \cite{Heemskerk:2009pn}. On the other hand, any single-trace operator that runs in one channel will induce its own correction to the double-trace data, as required by crossing \cite{Alday:2017gde}. The corrections can be systematically extracted using Caron-Huot's inversion formula \cite{Caron-Huot:2017vep}. To summarize this section, we write schematically a local four-point function as
\be
\braket{OOOO}= \mathbb{1}+\text{DT}^{(0)} + \frac{1}{N^2}\left(\text{ST}+\text{DT}^{(1)} \right) + \mathcal{O}(1/N^4) \label{OOOOdt}\,.
\ee
We will shortly see that when two of the operators are ANEC operators instead of local operators, the only contribution that survives the integral is the single-trace contribution. This will be a key point in what follows. Before we derive this fact, we start by reviewing the $N$-counting for operators whose two-point functions is not $\mathcal{O}(1)$ and we focus on the stress-tensor.

\subsubsection*{$N$-scaling for the stress-tensor}
The $N$-scaling for the stress-tensor is slightly different since
\be
\braket{TT}\sim N^2 \,, \qquad \braket{TTT}\sim N^2 \,, \qquad \braket{:T^2: :T^2:}\sim N^4 \,.
\ee
We therefore have the following scaling of the four-point function
\be
\braket{TTTT}= N^4(\mathbb{1}+\text{DT}^{(0)}) + N^2\left(\text{ST}+\text{DT}^{(1)} \right) + \mathcal{O}(1) \,.
\ee
The connected piece is still subleading compared to the disconnected piece, although the general scaling of the correlation function is different.

%\subsection{Three-point functions and the conformal collider bounds}
%We are now ready to start computing correlation functions, in particular to study the positivity property of the ANEC operator. For scalar states, we have already computed the three-point function in \rref{scalar3p} and positivity of the ANEC is trivial. For $U(1)$ current and stress-tensor states, we have from \rref{MTandJ}
%\bea \label{MJ}
%\braket{E}_J &=& \frac{q^0}{4\pi} \frac{1}{\eps_i^* \eps^i}  \eps^*_i \left[g^{ij} + a_2 \left(n^i n^j-\frac{1}{3}g^{ij}\right)\right] \eps_j \\
%\braket{E}_T &=&  \frac{q^0}{4\pi} \frac{1}{\eps_{ij}^* \eps^{ij}}\eps^*_{ij} \left[ g^{ik} g^{jl} + t_2 g^{ik}\left( n^j n^l -\frac{ g^{jl}}{3} \right) + t_4 \left(n^in^jn^kn^l- \frac{2g^{ik} g^{jl}}{15}  \right) \right] \eps_{kl} \notag
%\eea
%Positivity of these quantities for arbitrary polarizations yields the conformal collider bounds
%\bea
%-\frac{3}{2}&\leq& a_2  \ \ \leq 3 \notag \\
%0&\leq& 1-\frac{t_2}{3}-\frac{2t_4}{15} \notag \\
%0&\leq&t_2+ 2\left(1-\frac{t_2}{3}-\frac{2t_4}{15}\right) \notag \\
%0&\leq& t_2+t_4+ \frac{3}{2}\left(1-\frac{t_2}{3}-\frac{2t_4}{15}\right)  
%\eea
%derived in \cite{Hofman:2008ar}. The last three-inequalities can be recast to constrain the anomaly coefficients $a$ and $c$ as
%\be
% \frac{1}{3} \leq \frac{a}{c} \leq \frac{31}{18} \,,
%\ee 
%We now turn our attention to higher-point functions.
\subsection{Four-point functions}
\begin{figure}[h!]
\centering
\includegraphics[width=0.65\textwidth]{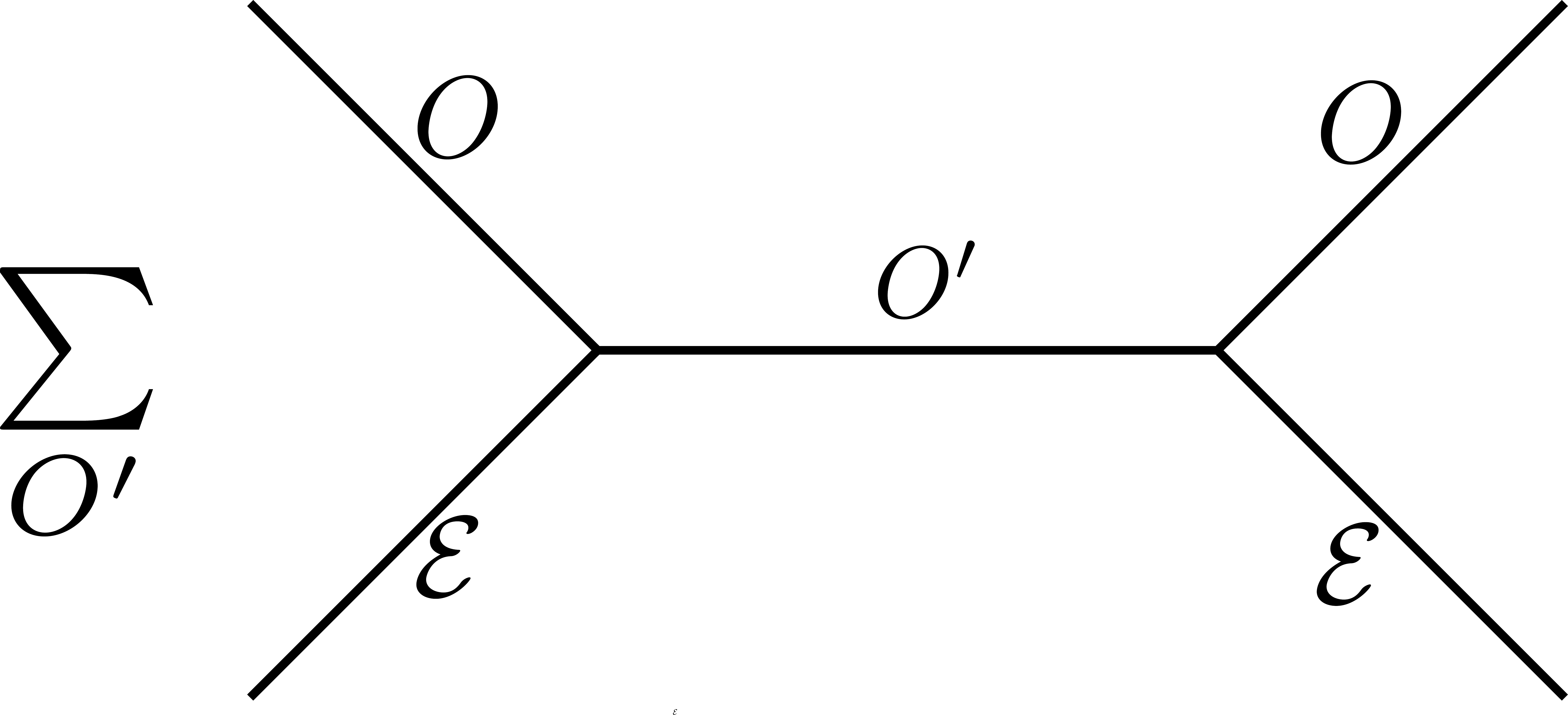}
\caption{The OPE expansion of the four-point function in the channel we have picked. To compute the correlator to the first two orders in $1/N^2$, the sum over $O'$ is over all single-trace and double-trace operators.}
\label{OPEpic}
\end{figure}

We would now like to compute four-point functions in large $N$ theories to first non-trivial order in the $1/N$ expansion. We can decompose a four-point function of two local operators and two ANECs using the OPE. We will always use the OPE channel where the local operator and the ANEC fuse, as illustrated in Fig. \ref{OPEpic} 

Following the discussion in section \ref{largeN}, the operators we need to take into account are all single-trace operators, as well as the double-trace operators and their corrected data at order $1/N^2$. We will start by showing that the contribution of all double-trace operators vanishes, both at leading level and when one takes into account their corrected data at order $1/N^2$.

\subsubsection{The fate of the double-trace operators}
In this section, we will show that the double-trace operators vanish in the four-point function of two ANECs and two local operators. To see this, we will consider the OPE of local operators
\be
T_{--} \times O \sim :T_{--} O: \,,
\ee
and integrate on both sides. At leading order in $N$, the three-point function (from which one could extract the OPE) is given by\footnote{If $O$ is the stress-tensor, there are other Wick contraction but they will vanish as well.}
\be \label{3pSSD}
\braket{T_{--}(x_1) O(x_2) :T_{--}O:(x_3)} \sim \braket{T_{--}(x_1)T_{--}(x_3)}\braket{O(x_2)O(x_3)} \,.
\ee
We now integrate on both sides. One can directly check that the integral vanishes
\be \label{int2p}
 \int \braket{T_{--}(x_1)T_{--}(x_3)} dx_1^- \sim \int dx_1^- \frac{(x_{12}^+)^4}{((x_{12}^{\perp})^2-x_{12}^-x_{12}^+)^6}=0 \,.
\ee
We thus conclude that double-trace operators $\text{DT}^{(0)}$  do not contribute at leading order in the $1/N$ expansion. Also, the identity is trivially projected out  by the same argument.

Therefore, we just need to discuss the corrections that appear at order $1/N^2$. Let us consider double trace operators in \rref{OOOOdt} denoted by DT$^{(1)}$.
%\bea
%\Delta_{n,l} &=& \Delta_{n,l}^{(0)} + \frac{1}{N^2}\gamma_{n,l}+... \notag \\
%C_{TT[OO]_{n,l}}&=&a^{(0)}_{n,l} +\frac{1}{N^2}a^{(1)}_{n,l}+... \,.
%\eea
We will now argue that these corrections vanish in the four-point function with two ANECs. To see this, first note that a change in the OPE coefficient would simply change the prefactor in \rref{3pSSD}, but the fact that it vanishes comes from the integral and a change in the overall coefficient is therefore irrelevant. The correction from the OPE coefficient thus doesn't contribute to the four-point function with two ANECs at this order.

The case of the anomalous dimensions is more subtle. If the operator acquires an anomalous dimension, the equations \rref{3pSSD} and \rref{int2p}  no longer hold, and the integral no longer vanishes. Instead, one can check that it picks up a piece proportional to $ \gamma_{n,l}/N^2$, which comes from the discontinuity of the integrand in \rref{3pSSD} once one includes the anomalous dimension. However, to compute the four-point function, one needs to do the OPE with both local operators, or in other words, to integrate twice. This means that the correction from the anomalous dimension to the double-trace operator will give a contribution of the order
\be
\frac{1}{N^4} \gamma_{n,l}^2 \,.
\ee
This is a direct consequence of the leading term $\text{DT}^{(0)}$ vanishing. If that was not the case we would indeed have corrections of order $1/N^2$. 

\begin{figure}[htb]
  	\begin{minipage}[t]{0.45\linewidth}
     			\centering \includegraphics[scale=0.3]{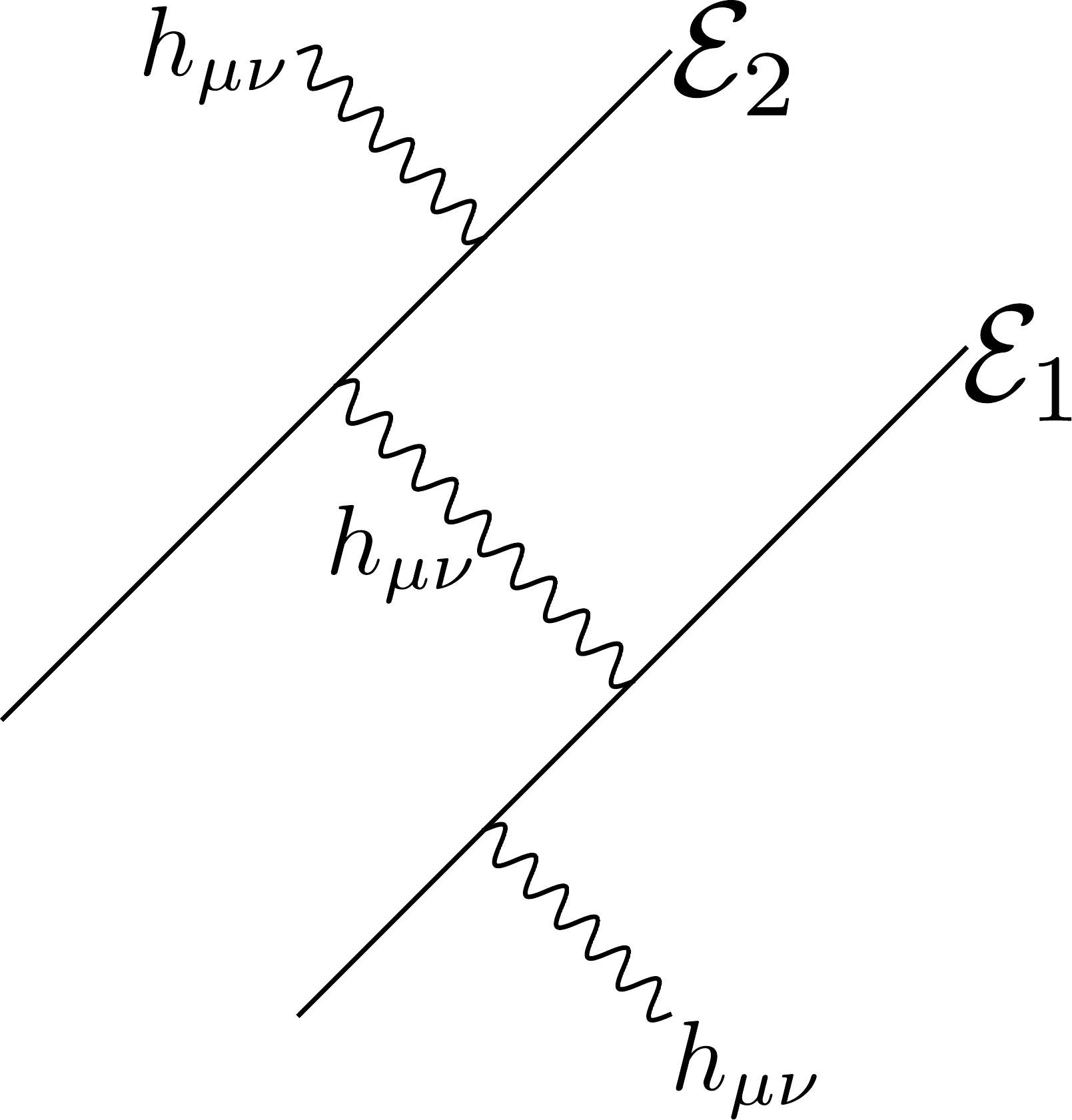}
      			\caption{A bulk picture of a graviton scattering through a shockwave. At tree-level the particle simply gets shifted when it passes through the shockwave and particle number is conserved.}
			\label{shockwaveTL}
   		\end{minipage}\hfill
   		\begin{minipage}[t]{0.45\linewidth}   
     			 \centering \includegraphics[scale=0.3]{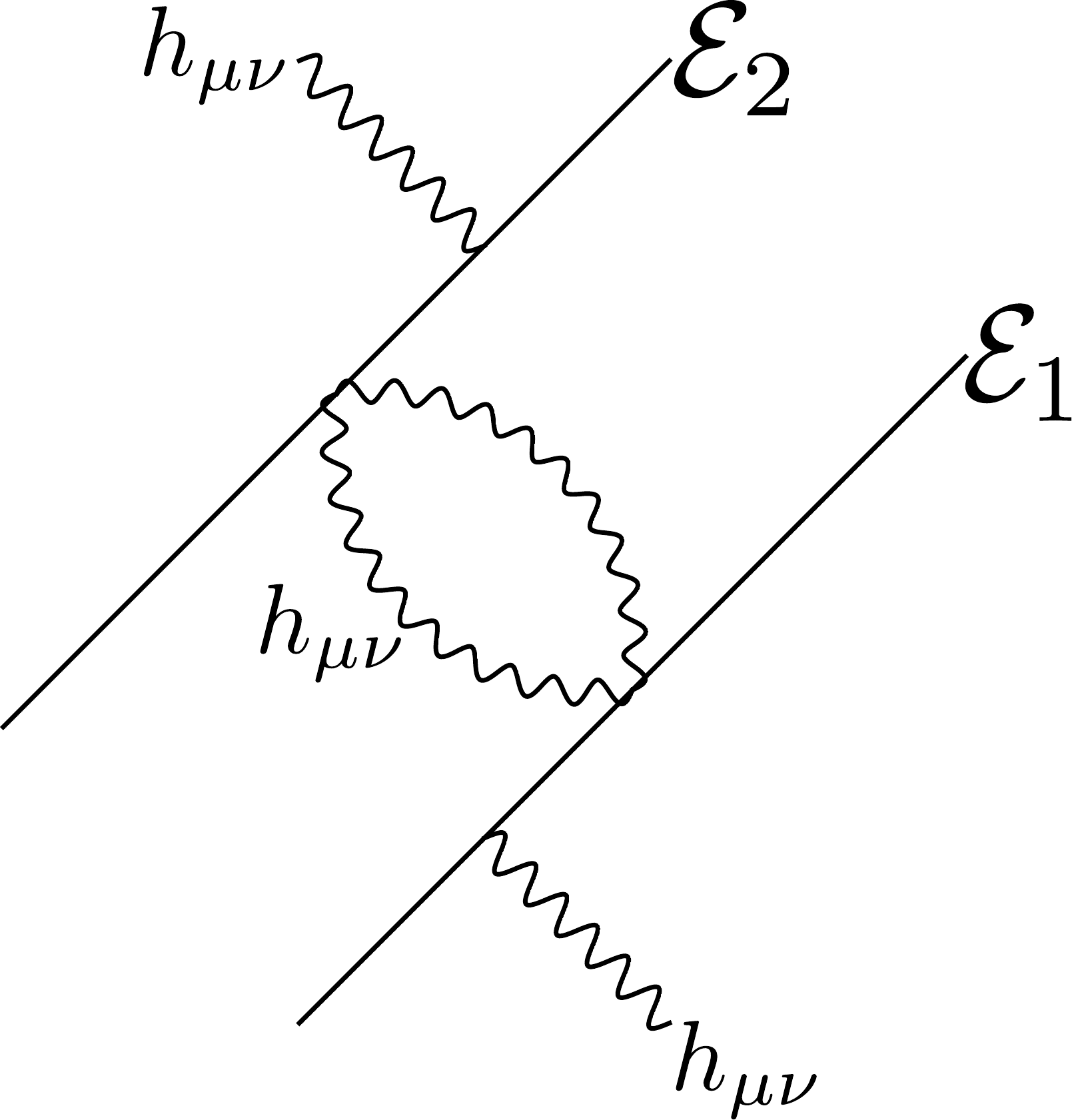}
     			 \caption{ A bulk picture of graviton production when passing through a shockwave. One can see that it is necessarily a loop effect.}
			\label{shockwave1L}
 	\end{minipage}
\end{figure}

The upshot is that $\text{DT}^{(1)}$  does not contribute at the order we are considering. It is worth mentioning that we are performing a computation at an order that corresponds to tree-level bulk physics. From the AdS point of view, the ANEC operators can be viewed as shockwaves and our computation should be thought of as propagating a particle through a shockwave \cite{Hofman:2008ar,Hofman:2009ug}. At tree-level, particle number must be conserved through the shockwave and all that happens to the particle is that it gets displaced. This is particularly clear for high energy particles that can only follow bulk geodesics. This is illustrated in Fig. \ref{shockwaveTL}. To consider the effect of particle creation, one would need to go one order higher in the $1/N$ expansion, which corresponds to loops in AdS. This is plotted in Fig. \ref{shockwave1L}. This is precisely the order $1/N^4$ and it is therefore not surprising that the effect of the double-trace operators can only be seen at that order.

We have now shown that the double-trace operators do not contribute at all to the order we care about, as advertised before. In two dimensions, it was true as an exact statement due to a symmetry. Here, it is valid thanks to the large $N$ limit, and only at this order. At higher orders, the double-trace operators would become important. Overall, we have shown that a four-point function will be given solely by the sum over single-trace operators ST. We now discuss their contribution.

\subsubsection{Single-trace operators and the effect of large gap}

\begin{figure}[h!]
\centering
\includegraphics[width=0.65\textwidth]{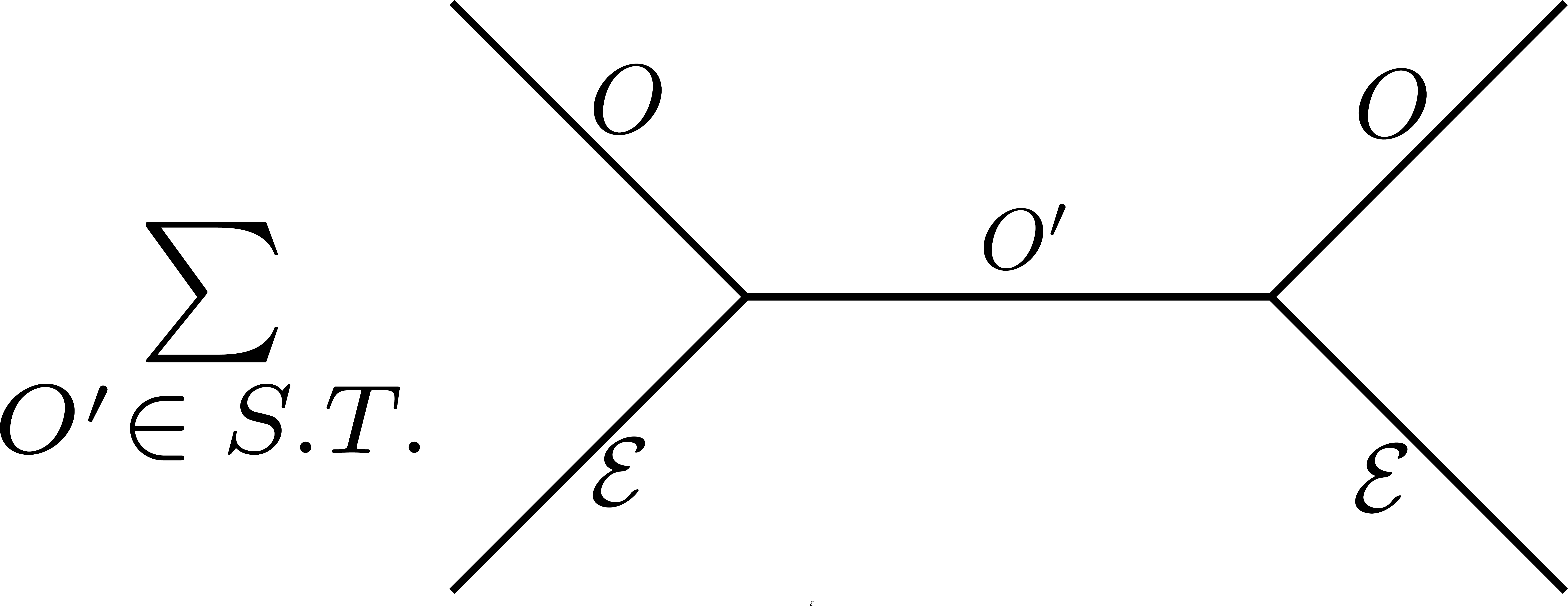}
\caption{The sum over operators has reduced to a sum over only the single-trace operators.}
\label{OPEpic2}
\end{figure}

We have shown that we only need to keep single-trace operators in the four-point function. This means that our calculation schematically reduces to Fig. \ref{OPEpic2}.

There is an operator that stands out in this sum. It is the operator that was used to create the state in the first place. When this particular operator is exchanged, the ANEC simply acts as a differential operator. To see this, note that at each three-point vertex, we have precisely the three-point function we computed exactly in section 3 using the differential operator. We therefore have
\be
\braket{O \EE_1\EE_2 O}_{O-\text{block}}= \mathcal{D}_1\mathcal{D}_2 \braket{OO} \,.
\ee
This turns out to be particularly simple for ANEC operators on the celestial sphere. In that setting, we showed that the differential operator becomes a transfer matrix between polarizations. We therefore have
\be
\frac{\bra{\mathbb{O}} E(n_1^i)E(n_2^i) \ket{\mathbb{O}}_{O-\text{block}} }{\braket{\mathbb{O}|\mathbb{O}}}= \left( \frac{q^0}{4\pi}\right)^2  \frac{1}{\epsilon^* \cdot \epsilon}\epsilon^* M_1M_2 \epsilon \,.
\ee
Thus, we simply multiply the transfer matrices. This is the whole advantage of thinking about the ANEC operator as a differential operator. Once we understand how it acts, if we have multiple ANECs we can simply apply one after the other even at finite distances. On the celestial sphere we just multiply the transfer matrices. It would also be straightforward to compute the all $O$ block to the $k$-point function of ANECs as well. It would be given by
\be
\frac{\bra{\mathbb{O}} E(n_1^i)...E(n_k^i) \ket{\mathbb{O}}_{\text{All} \  O}}{\braket{\mathbb{O}|\mathbb{O}}} = \left( \frac{q^0}{4\pi}\right)^k  \frac{1}{\epsilon^* \cdot \epsilon}\epsilon^* M_1...M_k \epsilon \,.
\ee
In a large $N$ CFT, this block would however not be enough. One would need to add to this the contribution of all other light single-trace operators. At this point, we will focus on theories with a large gap in the dimension of single-trace operators with spin $s>2$. In a large gap scenario, the higher spin single-trace operators are heavy and they do not contribute (or rather give small corrections suppressed by $1/\Delta_{\text{gap}}$ ). One could worry about other light operators of lower spin, but it was shown in \cite{Meltzer:2017rtf} that in a large gap scenario, couplings of these form are suppressed by the gap as well. We can, therefore, also neglect them. In a generic holographic CFT with no supersymmetry or other accidental symmetries we do not expect to have other low spin single trace operators, in any case.

We have thus arrived at the following conclusion:
\be
\notag \textbf{CFT with large N, large gap} \quad \Longrightarrow \quad \text{Only $O$ appears in the $\EE \ O$ OPE} \,.
\ee
The remainder of this paper will focus on drawing consequences or constraints from this statement. For example, we will see that the conformal collider bounds get squeezed in to give a definite value of $a/c=1$. In general, the OPE coefficients will be ``minimal" in that they match what Einstein gravity minimally coupled to matter would predict in the CFT.

\subsection{Einstein gravity from commutators}
In order to derive Einstein gravity and minimal couplings, we will study the commutator of two ANECs in the state created by a local operator. This amounts to computing a CFT four point function of the type described above. The commutator can be computed quite easily now that we know the transfer matrices. It is simply given by the commutator of the transfer matrices. 
As explained in the introduction, two ANEC operators must commute:
\be
[\EE_1,\EE_2]=0 \,.
\ee
For two ANEC operators on the same null-sheet (that means they are separated in the transverse direction) all points of the two null rays are space-like separated. Therefore, we expect  the operators to commute, see \cite{Casini:2017roe,Cordova:2018ygx}. When taken to the celestial sphere, one might worry about contributions to the commutator coming from the point at infinity. Given that in a CFT these observables can be related to the commutator above by a conformal transformation \cite{Hofman:2008ar}, we take this to be true even in this case.

We start by studying the commutator in states created by currents.
\subsubsection{$U(1)$ current states}
The commutator of the transfer matrices can be worked out from \rref{transferMatrix} and reads
\be
[M_1,M_2]^{i \, k} = \frac{3(8\tilde{e}-\tilde{c})}{2(\tilde{c}+\tilde{e})} \left(n_1 \cdot n_2 \right) \left( n_1^i n_2^k-n_2^in_1^k \right) \,.
\ee
Once again the tilde coefficients appearing above are related to the $JJT$ OPE and are defined in \cite{Osborn:1993cr}.
For this matrix to vanish for arbitrary polarization states, we must have
\be
\tilde{c}=8\tilde{e} \quad\quad \Longrightarrow \quad\quad a_2=0 \,.
\ee
This constraint on the OPE coefficients has a natural interpretation in the AdS dual. It corresponds to a minimal coupling between the bulk gauge field and the graviton. The effective action would be given by \cite{Hofman:2008ar}
\be
S\sim \int d^5x \sqrt{g} F_{\mu\nu}F^{\mu\nu} \,,
\ee
namely a Maxwell term. From a bulk effective field theory point of view, one could have written down non-minimal couplings involving curvature tensors, for example a coupling with the Weyl tensor $\int d^5x \sqrt{g}W_{\mu\nu\rho\sigma}F^{\mu\nu}F^{\rho\sigma}$,  but they would have modified the value of $\tilde{c}-8\tilde{e}$. We have therefore shown that the large gap assumption implies minimal couplings between gauge fields and the graviton.

\subsubsection{Stress-tensor states}
The transfer matrix for stress-tensor states is given from \rref{transferMatrix}, upon taking a suitable symmetrization and removing the traces. We find
\bea
M_{T}^{ij \, kl}&=&\frac{1}{2}(g^{ik}g^{jl}+g^{jk}g^{il})-\frac{1}{3}g^{ij}g^{kl}+t_2\Bigg[\frac{1}{4}(g^{ik}n^{j}n^{l}+g^{jk}n^{i}n^{l}+g^{il}n^{j}n^{k}+g^{jl}n^{i}n^{k}) \notag \\
&-&\frac{1}{3}(g^{ij}n^kn^l+n^in^jg^{kl})-\frac{1}{6}(g^{ik}g^{jl}+g^{jk}g^{il})+\frac{2}{9}g^{ij}g^{kl}\Bigg]  \\
&+&t_4\left[n^{i}n^{j} n^{k}n^{l}-\frac{1}{3}g^{ij}n^kn^l-\frac{1}{3}n^in^jg^{kl}-\frac{1}{15}(g^{ik}g^{jl}+g^{jk}g^{il})+\frac{7}{45}g^{ij}g^{kl} \right]\, . \notag
\eea
We now compute the commutator which reads 
\bea
[M_1,M_2]^{ij \, kl}&=&t_2^2\Bigg[\frac{n_1\cdot n_2}{8}(g^{ik}n_1^jn_2^l +g^{jk}n_1^in_2^l+g^{il}n_1^jn_2^k+g^{jl}n_1ijn_2^k)  \\
&-&\frac{1}{3}\left(n_1^in_1^jn_2^kn_2^l-\frac{1}{3}(g^{ij}n_2^kn_2^l+n_1^in_1^jg^{kl})\right)\Bigg] \notag \\
&+&t_4^2\left[ \left((n_1\cdot n_2)^2-\frac{1}{3}\right)\left(n_1^in_1^jn_2^kn_2^l-\frac{1}{3}(g^{ij}n_2^kn_2^l+n_1^in_1^jg^{kl})\right)\right] \notag \\
&-&t_2t_4\left[\frac{2}{3}\left(n_1^in_1^jn_2^kn_2^l-\frac{1}{3}(n_1^in_1^jg^{kl}+g^{ij}n_2^kn_2^l)\right)+\frac{(n_1\cdot n_2)^2}{3}\left(n_1^in_1^jg^{kl}+g^{ij}n_2^kn_2^l\right)\right] \notag \\
 &-& 1\leftrightarrow 2\,. \notag
\eea
Demanding that this vanishes when inserted in states of arbitrary polarizations yields 
\be
t_2=t_4=0 \,.
\ee
In terms of the anomaly coefficients, we have from \rref{aoverc}
\be
\frac{a}{c}=1 \,.
\ee
This corresponds to a bulk effective theory given by general relativity, without higher derivative corrections. More precisely, we have shown that all higher derivative corrections in \rref{eftgrav} are suppressed by a UV scale much larger than the IR scale $\Lambda$. In short,  in order for the commutator of the ANEC operators to vanish in a theory with a large gap, the bulk dual must be given by Einstein gravity.

\subsection{Strengthening of bounds from higher point correlators \label{betterbounds}}
There is an alternative route to these results. It is also possible to derive Einstein gravity with minimal couplings by considering higher-point functions of the ANEC operator.\footnote{D.H. would like to thank Sasha Zhiboedov for early discussions concerning this point. In particular for bringing up that the holographic computations of these quantities in the $AdS$ bulk show a similar phenomenon} The product of positive commuting operators must also be a positive operator. This implies
\be
\braket{E_1...E_k}\geq 0 \,.
\ee
We will now show that positivity of such correlators will strengthen the conformal collider bounds in this large $N$ scenario. Therefore, the window for non-minimal couplings will close in from both sides. To do so, we need to define a correlator that is ``blind" to the commuting properties of the ANEC operators. The most natural object to consider is the symmetrized correlator
\be
\braket{E_1...E_k}_{\text{SYM}} \equiv  \frac{1}{k!} \sum_{g\in S_k}\braket{E_{g(1)}...E_{g(k)}} \,.
\ee
One way to convince yourself that this is a reasonable observable is to think a bit about its holographic computation. This is discussed in \cite{Hofman:2008ar} for scalar states. The way to perform this computation is to push an incoming particle through a gravitational shockwave with insertions associated to each $E$ operator. When one performs the expansion of this solution the result is naturally symmetric under the reshuffling of all $E$'s as they all exist in the same light-like plane and have no natural ordering associated to them.

Consider, as an example, the observable above in a state created by a local current operator. We can now solve for the eigenvalues of such a matrix as a function of the parameter $a_2$ defined in \rref{OPEcouplings} using \rref{transferMatrix}. When one of the eigenvalues becomes zero, we are in danger of finding negative expectation values. The edges of the $a_2$ parameter space where the expectation values are positive are therefore given by the values of $a_2$ such that an eigenvalue vanishes for some angle on the celestial sphere.

\begin{figure}[h!]
\centering
\includegraphics[width=0.55\textwidth]{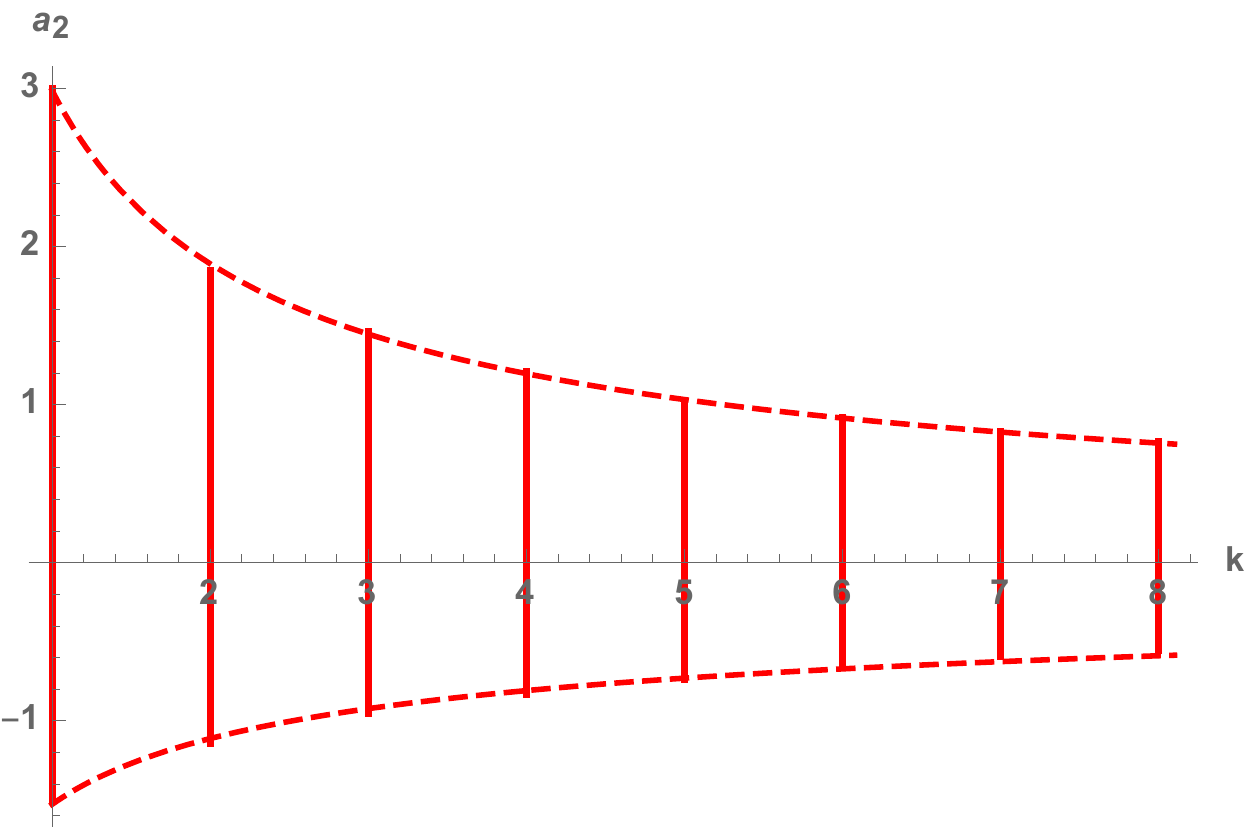}
\caption{A plot of the allowed parameter space for the coupling $a_2$, as demanded by the positivity of the $k$-point function of ANEC operators. For $k=1$, we have the conformal collider bounds. As we increase the number of operators, the region gets more and more constrained and is slowly closing in on $a_2=0$. We can fit the bounds by a power law and we find $\delta_{\max}\sim 2.98 k^{-0.66}$ and $\delta_{\min}\sim-1.53  k^{-0.46}$.}
\label{boundscurrent}
\end{figure}

We compute this numerically below. We distribute $k$ ANEC operators randomly over the celestial sphere and iterate the procedure many times to find the strongest possible constraint for a given $k$. We plot the results for current states in Fig. \ref{boundscurrent} for $k=1, \ldots , 8$. We clearly see that the bounds on $a_2$ close in on zero as we increase $k$. Demanding positivity of an arbitrary number of operator insertions will therefore close the allowed range down to $a_2=0$ which is again minimally coupled Maxwell theory in the bulk.

%\be
%M_{T}=\text{SYM}\left[g^{ik} g^{jl} + t_2 g^{ik}\left( n^j n^l -\frac{ g^{jl}}{3} \right) + t_4 \left(n^in^jn^kn^l- \frac{2g^{ik} g^{jl}}{15}  \right)\right] \,,
%\ee
%with
%\bea
%\text{SYM}\left[g^{ik} g^{jl}\right]&=&\frac{1}{2}(g^{ik}g^{jl}+g^{jk}g^{il})-\frac{1}{4}g^{ij}g^{kl} \notag \\
%\text{SYM}\left[g^{ik} n^{j}n^{l}\right]&=&\frac{1}{4}\left(g^{ik}n^{j}n^{l}+g^{jk}n^{i}n^{l}+g^{il}n^{j}n^{k}+g^{jl}n^{i}n^{k}-g^{ij}n^kn^l-n^in^jg^{kl}\right) \notag \\ 
%\text{SYM}\left[n^{i}n^{j} n^{k}n^{l}\right]&=&n^{i}n^{j} n^{k}n^{l}-\frac{1}{4}g^{ij}n^kn^l-\frac{1}{4}n^in^jg^{kl} 
%\eea
%We now compute the commutator which reads
%

\pagebreak

\section{Discussion}

In this paper, we have studied correlation functions of ANEC operators in states created by a local operator. We developed an OPE between the local operators and the ANEC operator and recast it as a differential operator. This statement is exact at the level of three-point functions and becomes approximately true for higher point functions in a CFT at large $N$ and with a large gap. The form of this differential operator is given as a series expansion which we were able to resum. In the limit where the ANEC operator is sent to the celestial sphere infinitely far away, the differential operator becomes particularly simple.

This formalism is particularly useful to compute correlation functions with multiple ANEC operators. In a large N CFT with a large gap to higher spin operators, we showed that the contribution of double-trace operators completely drops out from the correlator at the order that we care about and the correlation function of multiple ANEC operators is simply given by acting with a sequence of differential operators on the two-point function. The emerging structure is reminiscent of  $d=2$ physics, as it was previewed in section 2. 

We used this property to compute the commutator of two ANEC operators and demanded it must vanish. In a CFT with a large gap, we showed that this constrains the OPE coefficients of the theory to be ``minimal", which, in particular, forces the anomaly coefficients to satisfy
\be
a=c \,.
\ee
The bulk version of this statement is that any large $N$ theory with a large gap must have a holographic dual with Einstein gravity minimally coupled to matter. We have also computed the $k$-point function of ANEC operators and demanded it to be positive. This implies a strengthening of the conformal collider bounds. In the large $k$ limit, the bounds close in again on the minimal couplings. The two approaches turn out to be equivalent.

The most important direction in which this discussion could be improved concerns relaxing the assumption of an infinite gap to higher spin operators. If one kept a large but finite value of $\Delta_{gap}$, it would be possible to perform a systematic expansion in terms of this quantity in order to study how equalities like $a=c$ can be corrected by powers of $\Delta_{gap}^{-1}$. This way one could obtain precise expressions including numerical factors that would build on the results in \cite{Camanho:2014apa,Meltzer:2017rtf}. The obstacle is that this computation cannot be done reliably entirely in the conformal channel used in this paper where light-ray operators act on local operators creating a state. The reason is that it is actually easy to see that the addition of a finite number of operators to the computations described in this work cannot change the strong contraints coming from the vanishing of commutators. What we find in this case, instead, is the requirement that further non-minimal couplings to this new heavy operators must vanish as well. In order to have an effect on the constraints an infinite number of heavy operators need to be included. But this is equivalent to considering a finite number of them in the cross-channel. Therefore, our techniques don't apply directly to this case and they need important improvements to account for this physics.

If one could somehow improve the analysis, a new kind of results would become available. If many single trace operators can appear in the intermediate channel of the calculation of the ANEC commutator one would expect interesting sum rules to arise of the form\footnote{A similar type of structure was observed in \cite{Cordova:2017zej}.}

\be
\left(\frac{a-c}{c}\right)^2 \sim \sum_{O^{\Delta,s}_\text{S.T.} \neq T_{\mu\nu}} |C_{TTO}|^2 f(\Delta_O,s_O) \,.
\ee
One direct application that is readily available from the results presented here is the computation of arbitrary high-point correlation functions of ANEC operators, even when they are separated by a finite distance from each other in the large $N$ limit. While we have not looked at these observables in detail, it seems their structure is universal and can be thought of as a generalization of the well known structure present in two dimensional CFTs. The fact that we have written the operators in differential form gives us direct access to study the Ward identities for these theories, in the spirit of \cite{Belavin:1984vu}. We hope this approach will help provide a better understanding of the appearance of infinite dimensional algebras in some contexts in $d=4$ CFTs \cite{Beem:2013sza,Beem:2014kka,Cordova:2018ygx,Hofman:2018lfz}.

\section*{Acknowledgements}
It is a pleasure to thank Agnese Bissi, Alejandra Castro, Tom Hartman, Madalena Lemos, Onkar Parrikar, Jo\~ao Penedones, John Stout, Matt Walters and Sasha Zhiboedov for discussions. AB is supported by the NWO VENI grant 680-47-464 / 4114. D.H. and G.M. are supported in part by the ERC Starting Grant {\scriptsize{GENGEOHOL}}.

%
%
%
% and that operator becomes particularly simple in the limit of large distance where we push it to the celestial sphere. We can now easily compute higher point correlation functions, and for concreteness we will compute four-point functions\footnote{It is straightforward to generalize to higher-point functions.}. We will study scalar, current and stress-tensor states.
%
%
%\subsection{Scalar states}
%We start by considering scalar states. We study the four-point function
%\be
%\braket{O(x_4) \EE_2(x_2) \EE_1(x_1) O(x_3)} \,.
%\ee
%Given assumption 1, the correlation function evaluates to
%\be
%\braket{O(x_4) \EE_2(x_2) \EE_1(x_1) O(x_3)}=\mathcal{D}_2\mathcal{D}_1 \frac{1}{x_{34}^{2\D}} \,.
%\ee
%We will take the two-point function in momentum space with a four-momentum given by \rref{0k}. By rotational invariance, the first ANEC operator $\mathcal{E}_1$ can be set in the $z$ direction but the second operator can point anywhere. It is therefore given by
%\be
%\mathcal{D}_2\sim (2\pi i) \frac{1}{\pi^2}  \frac{1}{(\sP)^2} \frac{\p_{n^-}^2}{\p_{n^+}} \,,
%\ee
%where we defined
%\be
%n^{\pm}=t \pm n^i x^i \,,
%\ee
%with $n^i$ a three-vector pointing in the direction of the second operator $\mathcal{E}_2$. We quickly see that once we act on states of the form \rref{0k}, the operator $\mathcal{D}_2$ acts just like the first one. We therefore find
%\be
%\braket{E_1 E_2} = \frac{q^2}{16\pi^2} \,.
%\ee
%We see that the second moment of the energy density is trivial. In fact, they it is trivial to extend this to an arbitrary point function and we find
%\be
%\braket{E_1...E_n}=\frac{q^n}{(4\pi)^n}
%\ee 
%

\appendix

\section{Notation and Conventions \label{notation}}
We start by setting up some notation. We will work in $d=4$, and work mostly in lightcone coordinates
\be
x^\pm = t\pm z
\ee 
with metric
\be
ds^2=-dx^+dx^- + dx^2+dy^2 \,.
\ee
This also fixes the specification of the vectors $\xi_{\pm}^\mu$ from section \ref{opelocal}, and the vector $n^i$ would point in the $z$ direction, namely on the north pole of the celestial sphere. We will also use the more compact notation
\bea
\xperp&=&(x,y) \\
\xpsq&=&x^2+y^2
\eea

\section{$TO$ OPE \label{TOOPE}}
In this appendix, we want to compute the operator product expansion of the stress-tensor T when fusing with a scalar field $\mathcal{O}(x)$ of conformal weight $\Delta$. We obtain it by expanding the exact result \eqref{TOO} when the distance between two points is getting small. We follow \cite{Osborn:1993cr} and give more terms in the expansion. Also, we use the standard metric on $\mathbb{R}^4$, i.e $g_{\mu\nu}=\delta_{\mu\nu}$. First, let us define 
\be 
s^{\mu} = x^{\mu}-y^\mu,\qquad X^{\mu}=\frac{s^{\mu}}{s^2}-\frac{x^{\mu}-z^\mu}{(x-z)^2}.\label{sandX}
\ee
From \eqref{TOO} and \eqref{sandX}, the short distance limit of the three-point function in 4 dimensions is given by 
\be 
\langle T_{\mu\nu}(x)\mathcal{O}(y)\mathcal{O}(z)\rangle = \frac{1}{s^4(x-z)^4(y-z)^{2\Delta-4}}t_{\mu\nu}(X),\label{TOOOPE}
\ee
with $s$ and $X$ as in \eqref{sandX} and with $t_{\mu\nu}(X)$ given as in \eqref{tmunu}. The two-point function of two scalar operators is given by 
\be 
\langle \mathcal{O}(x)\mathcal{O}(y)\rangle =  \frac{1}{(x-y)^{2\Delta}}.
\ee
As $x\rightarrow y$, the three-point function can be expressed, using the OPE, as 

\begin{align} 
\langle T_{\mu\nu}(x)\mathcal{O}(y)\mathcal{O}(z)\rangle\sim& A_{\mu\nu}(s)\frac{1}{(y-z)^{2\Delta}}+B_{\mu\nu\alpha}(s)\frac{\partial}{\partial y^{\alpha}}\frac{1}{(x-y)^{2\Delta}}\\
&+C_{\mu\nu\alpha\beta}(s)\frac{\partial}{\partial y^{\alpha}}\frac{\partial}{\partial y^{\beta}}\frac{1}{(y-z)^{2\Delta}}+\dots\nonumber.
\end{align}
Expanding \eqref{TOOOPE} and matching order by order, one is able to determine the first coefficients $A_{\mu\nu}, B_{\mu\nu\alpha},\dots$. They can be built only out of two building blocks, namely the metric $\delta_{\mu\nu}$ and the vector $s_{\mu}$. The zeroth order term in $s$ is given by  
\be 
A_{\mu\nu}(s) = \frac{a}{s^4}\left(\frac{s_{\mu}s_{\nu}}{s^2}-\frac{1}{4}\delta_{\mu\nu}\right)= \frac{a}{s^4}t_{\mu\nu}(s).
\ee
The first order term in $s$ is given by 
\be 
B_{\mu\nu\alpha}(s) = \frac{a}{2\Delta s^4}\left(s_{\mu}\delta_{\alpha\nu}+s_{\nu}\delta_{\alpha\mu}-s_{\alpha}\delta_{\mu\nu}+4\frac{s_{\mu}s_{\nu}s_{\alpha}}{s^2}\right).
\ee
The second order term in $s$ is given by
\begin{align}
C_{\mu\nu\alpha\beta}(s) &=C_1 s^2 \delta _{\mu  \nu }\delta _{\beta  \alpha } +C_2s^2 \left(\delta _{\beta  \nu } \delta _{\mu  \alpha }+\delta _{\beta  \mu } \delta _{\nu  \alpha }\right)+C_3(s_{\beta } s_{\alpha } \delta _{\mu  \nu }+s_{\mu } s_{\nu } \delta _{\beta  \alpha })\\
&+C_4(s_{\nu } s_{\alpha } \delta _{\beta  \mu }+s_{\mu } s_{\alpha } \delta _{\beta  \nu }+s_{\beta } s_{\nu } \delta _{\mu 
   \alpha }+s_{\beta } s_{\mu } \delta _{\nu  \alpha })+C_5\frac{s_{\beta } s_{\mu } s_{\nu } s_{\alpha }}{s^2}\,,\nonumber
\end{align} 
with 
\begin{align}
C_1 {}=& \frac{a}{8 \left(\Delta ^2+\Delta \right) s^4},\\
C_2 =& \frac{a}{8 \left(\Delta ^2+\Delta \right) s^4},\\
C_3=&-\frac{3 a}{4 \left(\Delta ^2+\Delta \right) s^4},\\
C_4=&\frac{a}{2 \left(\Delta ^2+\Delta \right) s^4},\\
C_5=& \frac{a}{\left(\Delta ^2+\Delta \right) s^4}.
\end{align}
The third order term in $s$ is given by
\begin{align} 
D_{\mu\nu\alpha\beta\chi}(s) &=D_1 s^2 \left(s_{\nu } \delta _{\alpha  \chi } \delta _{\beta  \mu }+s_{\mu } \delta _{\alpha  \chi } \delta _{\beta  \nu }+s_{\nu } \delta _{\alpha  \mu } \delta _{\beta  \chi }+s_{\mu } \delta _{\alpha  \nu } \delta _{\beta  \chi }+s_{\nu } \delta _{\alpha 
   \beta } \delta _{\mu  \chi }+s_{\mu } \delta _{\alpha  \beta } \delta _{\nu  \chi }\right)\nonumber\\
  & + D_2s^2 \left(s_{\chi } \delta _{\alpha  \beta } \delta _{\mu  \nu }+s_{\beta } \delta _{\alpha  \chi } \delta _{\mu  \nu }+s_{\alpha } \delta _{\beta  \chi } \delta _{\mu  \nu }\right)\nonumber\\  
  & + D_3s^2 \left(s_{\chi } \delta _{\alpha  \nu } \delta _{\beta  \mu }+s_{\beta } \delta _{\alpha  \nu } \delta _{\mu  \chi }+s_{\chi } \delta _{\alpha  \mu } \delta _{\beta  \nu }+s_{\alpha } \delta _{\beta  \nu } \delta _{\mu  \chi }+s_{\beta } \delta
   _{\alpha  \mu } \delta _{\nu  \chi }+s_{\alpha } \delta _{\beta  \mu } \delta _{\nu  \chi }\right)\nonumber\\
   &+D_4(s_{\mu } s_{\nu } s_{\chi } \delta _{\alpha  \beta }+s_{\beta } s_{\mu } s_{\nu } \delta _{\alpha  \chi }+s_{\alpha } s_{\mu } s_{\nu } \delta _{\beta  \chi })\nonumber\\
   & + D_5(s_{\beta } s_{\nu } s_{\chi } \delta _{\alpha  \mu }+s_{\alpha } s_{\nu } s_{\chi } \delta _{\beta  \mu }+s_{\beta } s_{\mu } s_{\chi } \delta _{\alpha  \nu }+s_{\alpha } s_{\mu } s_{\chi } \delta _{\beta  \nu }+s_{\alpha } s_{\beta } s_{\nu } \delta
   _{\mu  \chi }+s_{\alpha } s_{\beta } s_{\mu } \delta _{\nu  \chi })\nonumber\\
   &+ D_6s_{\alpha } s_{\beta } s_{\chi } \delta _{\mu  \nu }+D_7\frac{s_{\alpha } s_{\beta } s_{\mu } s_{\nu } s_{\chi }}{s^2},
\end{align}
where 
\begin{align}
D_1&= -\frac{a}{8 \Delta  (\Delta +1) (\Delta +2) s^4},\\
D_2 &= \frac{a}{8 \Delta  (\Delta +1) (\Delta +2) s^4},\\
D_3&=\frac{a}{8 \Delta  (\Delta +1) (\Delta +2) s^4},\\
D_4&=-\frac{a}{2 \Delta  (\Delta +1) (\Delta +2) s^4},\\
D_5 &=\frac{a}{2 \Delta  (\Delta +1) (\Delta +2) s^4},\\
D_6&= -\frac{a}{\Delta  (\Delta +1) (\Delta +2) s^4},\\
D_7 &= \frac{a}{\Delta  (\Delta +1) (\Delta +2) s^4}.
\end{align}
The fourth order term is the last one we will write explicitly, and it is given by
\begin{align}
E_{\mu\nu\alpha\beta\chi\delta} &= E_1 s^4 \left(\delta _{\beta  \delta } \delta _{\alpha  \chi } \delta _{\mu  \nu }+\delta _{\alpha  \delta } \delta _{\beta
    \chi } \delta _{\mu  \nu }+\delta _{\delta  \chi } \delta _{\alpha  \beta } \delta _{\mu  \nu }\right)\nonumber\\
    & + E_2s^2 \left(s_{\delta } s_{\chi } \delta _{\alpha  \beta } \delta _{\mu  \nu }+\delta _{\alpha  \delta } s_{\beta }
   s_{\chi } \delta _{\mu  \nu }+\delta _{\beta  \delta } s_{\alpha } s_{\chi } \delta _{\mu  \nu }+s_{\beta }
   s_{\delta } \delta _{\alpha  \chi } \delta _{\mu  \nu }+s_{\alpha } s_{\delta } \delta _{\beta  \chi } \delta _{\mu 
   \nu }+\delta _{\delta  \chi } s_{\alpha } s_{\beta } \delta _{\mu  \nu }\right)\nonumber\\
&+E_3s^4 \left(\delta _{\delta  \mu } \delta _{\alpha  \chi } \delta _{\beta  \nu }+\delta _{\delta  \nu } \delta _{\alpha 
   \chi } \delta _{\beta  \mu }+\delta _{\delta  \mu } \delta _{\alpha  \nu } \delta _{\beta  \chi }+\delta _{\delta 
   \nu } \delta _{\alpha  \mu } \delta _{\beta  \chi }+\delta _{\delta  \chi } \delta _{\alpha  \nu } \delta _{\beta 
   \mu }+\delta _{\delta  \chi } \delta _{\alpha  \mu } \delta _{\beta  \nu }\right.\nonumber\\
   &\left.+\delta _{\beta  \delta } \delta _{\alpha 
   \nu } \delta _{\mu  \chi }+\delta _{\alpha  \delta } \delta _{\beta  \nu } \delta _{\mu  \chi }+\delta _{\delta  \nu
   } \delta _{\alpha  \beta } \delta _{\mu  \chi }+\delta _{\beta  \delta } \delta _{\alpha  \mu } \delta _{\nu  \chi
   }+\delta _{\alpha  \delta } \delta _{\beta  \mu } \delta _{\nu  \chi }+\delta _{\delta  \mu } \delta _{\alpha  \beta
   } \delta _{\nu  \chi }\right)\nonumber\\
   & + E_4s^2 \left(s_{\delta } s_{\chi } \delta _{\alpha  \nu } \delta _{\beta  \mu }+\delta _{\delta  \mu } s_{\beta } s_{\chi
   } \delta _{\alpha  \nu }+s_{\beta } s_{\delta } \delta _{\alpha  \nu } \delta _{\mu  \chi }+s_{\delta } s_{\chi }
   \delta _{\alpha  \mu } \delta _{\beta  \nu }+\delta _{\delta  \mu } s_{\alpha } s_{\chi } \delta _{\beta  \nu
   }\right.\nonumber\\&+\delta _{\delta  \nu } s_{\beta } s_{\chi } \delta _{\alpha  \mu }+\delta _{\delta  \nu } s_{\alpha } s_{\chi }
   \delta _{\beta  \mu }
   \left.+s_{\alpha } s_{\delta } \delta _{\beta  \nu } \delta _{\mu  \chi }+\delta _{\delta  \nu }
   s_{\alpha } s_{\beta } \delta _{\mu  \chi }+s_{\beta } s_{\delta } \delta _{\alpha  \mu } \delta _{\nu  \chi
   }+s_{\alpha } s_{\delta } \delta _{\beta  \mu } \delta _{\nu  \chi }+\delta _{\delta  \mu } s_{\alpha } s_{\beta }
   \delta _{\nu  \chi }\right)\nonumber\\
   &+E_5(s_{\beta } s_{\delta } s_{\nu } s_{\chi } \delta _{\alpha  \mu }+s_{\alpha } s_{\delta } s_{\nu } s_{\chi } \delta
   _{\beta  \mu }+\delta _{\delta  \mu } s_{\alpha } s_{\beta } s_{\nu } s_{\chi }+s_{\beta } s_{\delta } s_{\mu }
   s_{\chi } \delta _{\alpha  \nu }\nonumber\\
   &+s_{\alpha } s_{\delta } s_{\mu } s_{\chi } \delta _{\beta  \nu }+\delta _{\delta 
   \nu } s_{\alpha } s_{\beta } s_{\mu } s_{\chi }+s_{\alpha } s_{\beta } s_{\delta } s_{\nu } \delta _{\mu  \chi
   }+s_{\alpha } s_{\beta } s_{\delta } s_{\mu } \delta _{\nu  \chi })\nonumber\\
   &+ E_6s^2 \left(\delta _{\beta  \delta } s_{\nu } s_{\chi } \delta _{\alpha  \mu }+s_{\delta } s_{\nu } \delta _{\alpha  \mu
   } \delta _{\beta  \chi }+\delta _{\delta  \chi } s_{\beta } s_{\nu } \delta _{\alpha  \mu }+\delta _{\alpha  \delta
   } s_{\nu } s_{\chi } \delta _{\beta  \mu }+\delta _{\delta  \mu } s_{\nu } s_{\chi } \delta _{\alpha  \beta }+\delta
   _{\beta  \delta } s_{\mu } s_{\chi } \delta _{\alpha  \nu }\right.\nonumber\\
   &\left.+\delta _{\alpha  \delta } s_{\mu } s_{\chi } \delta
   _{\beta  \nu }+\delta _{\delta  \nu } s_{\mu } s_{\chi } \delta _{\alpha  \beta }+s_{\delta } s_{\nu } \delta
   _{\alpha  \chi } \delta _{\beta  \mu }+\delta _{\delta  \mu } s_{\beta } s_{\nu } \delta _{\alpha  \chi }+s_{\delta
   } s_{\mu } \delta _{\alpha  \chi } \delta _{\beta  \nu }+\delta _{\delta  \nu } s_{\beta } s_{\mu } \delta _{\alpha 
   \chi }\right.\nonumber\\
   &\left.+\delta _{\delta  \mu } s_{\alpha } s_{\nu } \delta _{\beta  \chi }+s_{\delta } s_{\mu } \delta _{\alpha  \nu
   } \delta _{\beta  \chi }+\delta _{\delta  \nu } s_{\alpha } s_{\mu } \delta _{\beta  \chi }+\delta _{\delta  \chi }
   s_{\alpha } s_{\nu } \delta _{\beta  \mu }+\delta _{\delta  \chi } s_{\beta } s_{\mu } \delta _{\alpha  \nu }+\delta
   _{\delta  \chi } s_{\alpha } s_{\mu } \delta _{\beta  \nu }\right.\nonumber\\
   &\left.+s_{\delta } s_{\nu } \delta _{\alpha  \beta } \delta
   _{\mu  \chi }+\delta _{\alpha  \delta } s_{\beta } s_{\nu } \delta _{\mu  \chi }+\delta _{\beta  \delta } s_{\alpha
   } s_{\nu } \delta _{\mu  \chi }+s_{\delta } s_{\mu } \delta _{\alpha  \beta } \delta _{\nu  \chi }+\delta _{\alpha 
   \delta } s_{\beta } s_{\mu } \delta _{\nu  \chi }+\delta _{\beta  \delta } s_{\alpha } s_{\mu } \delta _{\nu  \chi
   }\right)\nonumber\\
   & + E_7(s_{\delta } s_{\mu } s_{\nu } s_{\chi } \delta _{\alpha  \beta }+\delta _{\alpha  \delta } s_{\beta } s_{\mu } s_{\nu }
   s_{\chi }+\delta _{\beta  \delta } s_{\alpha } s_{\mu } s_{\nu } s_{\chi }+s_{\beta } s_{\delta } s_{\mu } s_{\nu }
   \delta _{\alpha  \chi }+s_{\alpha } s_{\delta } s_{\mu } s_{\nu } \delta _{\beta  \chi }+\delta _{\delta  \chi }
   s_{\alpha } s_{\beta } s_{\mu } s_{\nu })\nonumber\\
   & + E_8s^2 \left(\delta _{\beta  \delta } s_{\mu } s_{\nu } \delta _{\alpha  \chi }+\delta _{\alpha  \delta } s_{\mu } s_{\nu
   } \delta _{\beta  \chi }+\delta _{\delta  \chi } s_{\mu } s_{\nu } \delta _{\alpha  \beta }\right)\nonumber\\
   &+ E_9 s_{\alpha } s_{\beta } s_{\delta } s_{\chi } \delta _{\mu  \nu }+E_{10}\frac{s_{\alpha } s_{\beta } s_{\delta } s_{\mu } s_{\nu } s_{\chi }}{s^2},
    \end{align}
    with 
    \begin{align}
    E_1=& -\frac{a}{64 \Delta  (\Delta +1) (\Delta +2) (\Delta +3) s^4}\,,\\
    E_2 =& \frac{a}{8 \Delta  (\Delta +1) (\Delta +2) (\Delta +3) s^4}\,,\\
    E_3 =&-\frac{a}{64 \Delta  (\Delta +1) (\Delta +2) (\Delta +3) s^4}\,, \\
    E_4 =& \frac{a}{8 \Delta  (\Delta +1) (\Delta +2) (\Delta +3) s^4}\,,\\
    E_5 =& \frac{a}{2 \Delta  (\Delta +1) (\Delta +2) (\Delta +3) s^4}\,,\\
    E_6 =& -\frac{3 a}{32 \Delta  (\Delta +1) (\Delta +2) (\Delta +3) s^4}\,,\\
    E_7=&-\frac{3 a}{8 \Delta  (\Delta +1) (\Delta +2) (\Delta +3) s^4} \,,\\
    E_8=& \frac{a}{8 \Delta  (\Delta +1) (\Delta +2) (\Delta +3) s^4}\,,\\
     E_9 =&-\frac{5 a}{4 \Delta  (\Delta +1) (\Delta +2) (\Delta +3) s^4}\,, \\
    E_{10}=& \frac{a}{\Delta  (\Delta +1) (\Delta +2) (\Delta +3) s^4}.
        \end{align}
To compute the OPE with the ANEC operator. We can integrate these terms order by order in the OPE expansion. The first two vanish upon integration and $C_{\mu\nu\rho\sigma}$ is the first order that contributes. Performing the integrals and setting $s^{\perp}=0$ gives the differential operator \rref{Dseries}.

\section{Differential operator acting on $T$ and $J$ \label{diffopspin}}

\subsection{$U(1)$ currents}

In this section, we explain the structure of the OPE between the ANEC operator and conserved currents. Following the rules established in section \ref{6rules}, the most general differential operator that one can write down is of the following form
\bea \label{diffopgencur}
\EE \epsilon^\mu J_\mu&=&  \left(\sum_{q_1,r_1,s_1,t_1} c^{1}_{q_1,r_1,s_1,t_1}\left(x_{12}  \cdot \xi_-\right)^{q_1} \left(\p \cdot \xi_+\right)^{r_1} \left(\p \cdot \xi_-\right)^{s_1}  \left(\p \cdot \p\right)^{t_1}\right)\epsilon \cdot J  \\
&+& \left(\sum_{q_2,r_2,s_2,t_2} c^{2}_{q_2,r_2,s_2,t_2}\left(x_{12}  \cdot \xi_-\right)^{q_2} \left(\p \cdot \xi_+\right)^{r_2} \left(\p \cdot \xi_-\right)^{s_2}  \left(\p \cdot \p\right)^{t_2}\right)(\epsilon \cdot \xi_+)(\xi_+ \cdot J) \notag \\
&+& \left(\sum_{q_3,r_3,s_3,t_3} c^{3}_{q_3,r_3,s_3,t_3}\left(x_{12}  \cdot \xi_-\right)^{q_3} \left(\p \cdot \xi_+\right)^{r_3} \left(\p \cdot \xi_-\right)^{s_3}  \left(\p \cdot \p\right)^{t_3}\right)(\epsilon \cdot \xi_+)(\xi_- \cdot J) \notag \\
&+& \left(\sum_{q_4,r_4,s_4,t_4} c^{4}_{q_4,r_4,s_4,t_4}\left(x_{12}  \cdot \xi_-\right)^{q_4} \left(\p \cdot \xi_+\right)^{r_4} \left(\p \cdot \xi_-\right)^{s_4}  \left(\p \cdot \p\right)^{t_4}\right)(\epsilon \cdot \xi_-)(\xi_+ \cdot J) \notag \\
&+& \left(\sum_{q_5,r_5,s_5,t_5} c^{5}_{q_5,r_5,s_5,t_5}\left(x_{12}  \cdot \xi_-\right)^{q_5} \left(\p \cdot \xi_+\right)^{r_5} \left(\p \cdot \xi_-\right)^{s_5}  \left(\p \cdot \p\right)^{t_5}\right)(\epsilon \cdot \xi_-)(\xi_- \cdot J) \notag \\
&+& \left(\sum_{q_6,r_6,s_6,t_6} c^{6}_{q_6,r_6,s_6,t_6}\left(x_{12}  \cdot \xi_-\right)^{q_6} \left(\p \cdot \xi_+\right)^{r_6} \left(\p \cdot \xi_-\right)^{s_6}  \left(\p \cdot \p\right)^{t_6}\right)(\epsilon \cdot \p)(\xi_+ \cdot J) \notag \\
&+& \left(\sum_{q_7,r_7,s_7,t_7} c^{7}_{q_7,r_7,s_7,t_7}\left(x_{12}  \cdot \xi_-\right)^{q_7} \left(\p \cdot \xi_+\right)^{r_7} \left(\p \cdot \xi_-\right)^{s_7}  \left(\p \cdot \p\right)^{t_7}\right)(\epsilon \cdot \p)(\xi_- \cdot J) \notag \,,
\eea
with the conditions
\bea
-q_1+r_1+s_1+2t_1 &=&3 \,, \qquad q_1-r_1+s_1=1 \notag \\
-q_2+r_2+s_2+2t_2 &=&3 \,, \qquad q_2-r_2+s_2=3 \notag \\
-q_3+r_3+s_3+2t_3 &=&3 \,, \qquad q_3-r_3+s_3=1 \notag \\
-q_4+r_4+s_4+2t_4 &=&3 \,, \qquad q_4-r_4+s_4=1  \\
-q_5+r_5+s_5+2t_5 &=&3 \,, \qquad q_5-r_5+s_5=-1 \notag \\
-q_6+r_6+s_6+2t_6 &=&2 \,, \qquad q_6-r_6+s_6=2 \notag \\
-q_7+r_7+s_7+2t_7 &=&2 \,, \qquad q_7-r_7+s_7=0 \notag
\eea
Note that due to conservation of the current, we have not allowed contraction between $\p^\mu$ and $J^\mu$, since it vanishes. We can compute the values of the coefficients $c^i$ at any given order by expanding the integrated three-point function, and find a similar expression to \rref{Dseries}. It is then easy to resum the operator and look at the large distance limit. When acting on zero-momentum eigenstates, the operator is
\bea
E \epsilon^\mu J_\mu &=& \frac{-\pi q^0}{4}\left( \frac{18\tilde{e}}{c_v} \epsilon_-J_+ + 3\frac{\tilde{c}-2\tilde{e}}{c_v} \epsilon_+J_- +3\frac{\tilde{c}-8\tilde{e}}{c_v} \epsilon_+ J_+ -\frac{3(\tilde{c}-2\tilde{e})}{2c_v} \epsilon_iJ_i \right) \notag \\
&=& \frac{\pi q^0}{4}\left(\frac{3(\tilde{c}-2\tilde{e})}{2c_v}\epsilon \cdot J -3\frac{\tilde{c}-8\tilde{e}}{c_v}\left(\xi_+ \cdot J\right)( \xi_+-\xi_-) \cdot \epsilon \right)  \,.
\eea
%\bea
%E \epsilon^\mu J_\mu &=& \frac{-iq}{2}\left( \frac{18\tilde{e}}{c_v} \epsilon_-J_+ + 3\frac{\tilde{c}-2\tilde{e}}{c_v} \epsilon_+J_- +3\frac{\tilde{c}-8\tilde{e}}{c_v} \epsilon_+ J_+ -\frac{3(\tilde{c}-2\tilde{e})}{2c_v} \epsilon_iJ_i \right) \notag \\
%&=& \frac{-iq}{2}\left(-\frac{3(\tilde{c}-2\tilde{e})}{2c_v}\epsilon \cdot J +3\frac{\tilde{c}-8\tilde{e}}{c_v}\left(\xi_+ \cdot J\right)( \xi_+-\xi_-) \cdot \epsilon \right)  \,.
%\eea
$c_v$ is the coefficient appearing in the two-point function and reads
\be
c_v=\pi^2(\tilde{c}+\tilde{e})
\ee

\subsection{The stress-tensor}
The most general operator that one can write down for the stress-tensor is
{\allowdisplaybreaks
\begin{align} \label{diffopgenT} 
\EE \epsilon^{\mu\nu} T_{\mu\nu} =&\difop{1}\epsilon_{\mu\nu} T^{\mu\nu}\\  \notag
 + & \difop{2}( \epsilon_{\mu\nu}\xi_{+}^{\mu}\xi_{+}^{\nu})(\xi_{+}^{\alpha}\xi_{+}^{\beta}  T_{\alpha\beta})\\  \notag
  + & \difop{3}( \epsilon_{\mu\nu}\xi_{+}^{\mu}\xi_{+}^{\nu})(\xi_{+}^{\alpha}\xi_{-}^{\beta}  T_{\alpha\beta})\\  \notag
   + & \difop{4}( \epsilon_{\mu\nu}\xi_{+}^{\mu}\xi_{+}^{\nu})(\xi_{-}^{\alpha}\xi_{-}^{\beta}  T_{\alpha\beta})\\ \notag
    + & \difop{5}( \epsilon_{\mu\nu}\xi_{+}^{\mu}\xi_{-}^{\nu})(\xi_{+}^{\alpha}\xi_{+}^{\beta}  T_{\alpha\beta})\\ \notag
    + & \difop{6}( \epsilon_{\mu\nu}\xi_{+}^{\mu}\xi_{-}^{\nu})(\xi_{+}^{\alpha}\xi_{-}^{\beta}  T_{\alpha\beta})\\ \notag
     + & \difop{7}( \epsilon_{\mu\nu}\xi_{+}^{\mu}\xi_{-}^{\nu})(\xi_{-}^{\alpha}\xi_{-}^{\beta}  T_{\alpha\beta})\\   \notag
      + & \difop{8}( \epsilon_{\mu\nu}\xi_{-}^{\mu}\xi_{-}^{\nu})(\xi_{+}^{\alpha}\xi_{+}^{\beta}  T_{\alpha\beta})\\  \notag
      + & \difop{9}( \epsilon_{\mu\nu}\xi_{-}^{\mu}\xi_{-}^{\nu})(\xi_{+}^{\alpha}\xi_{-}^{\beta}  T_{\alpha\beta})\\   \notag
      + & \difop{10}( \epsilon_{\mu\nu}\xi_{-}^{\mu}\xi_{-}^{\nu})(\xi_{-}^{\alpha}\xi_{-}^{\beta}  T_{\alpha\beta})\\ \notag
       + & \difop{11}( \epsilon_{\mu\nu}\xi_{+}^{\mu}\partial^{\nu})(\xi_{+}^{\alpha}\xi_{+}^{\beta} T_{\alpha\beta})\\ \notag
        + & \difop{12}( \epsilon_{\mu\nu}\xi_{+}^{\mu}\partial^{\nu})(\xi_{+}^{\alpha}\xi_{-}^{\beta}T_{\alpha\beta})\\ \notag
         + & \difop{13}( \epsilon_{\mu\nu}\xi_{+}^{\mu}\partial^{\nu})(\xi_{-}^{\alpha}\xi_{-}^{\beta}T_{\alpha\beta})\\ \notag
             + & \difop{14}( \epsilon_{\mu\nu}\xi_{-}^{\mu}\partial^{\nu})(\xi_{+}^{\alpha}\xi_{+}^{\beta}T_{\alpha\beta})\\ \notag
           + & \difop{15}( \epsilon_{\mu\nu}\xi_{-}^{\mu}\partial^{\nu})(\xi_{+}^{\alpha}\xi_{-}^{\beta}T_{\alpha\beta})\\ \notag
            + & \difop{16}( \epsilon_{\mu\nu}\xi_{-}^{\mu}\partial^{\nu})(\xi_{-}^{\alpha}\xi_{-}^{\beta}T_{\alpha\beta})\\ \notag
             + & \difop{17}(\epsilon_{\mu\nu}\partial^{\mu}\partial^{\nu})(\xi_{+}^{\alpha}\xi_{+}^{\beta}T_{\alpha\beta})\\ \notag
              + & \difop{18}(\epsilon_{\mu\nu}\partial^{\mu}\partial^{\nu})(\xi_{+}^{\alpha}\xi_{-}^{\beta}T_{\alpha\beta})\\ \notag
              + & \difop{19}(\epsilon_{\mu\nu}\partial^{\mu}\partial^{\nu})(\xi_{-}^{\alpha}\xi_{-}^{\beta}T_{\alpha\beta})\\ \notag
                + & \difop{20}( \xi_{+\mu}(\epsilon^{\mu\gamma} T_{\gamma\nu})\xi_{+}^{\nu})\\ \notag
                  + & \difop{21}( \xi_{+\mu}(\epsilon^{\mu\gamma} T_{\gamma\nu})\xi_{-}^{\nu})\\ \notag
                    + & \difop{22}( \xi_{-\mu}(\epsilon^{\mu\gamma} T_{\gamma\nu})\xi_{+}^{\nu})\\ \notag
                      + & \difop{23}( \xi_{-\mu}(\epsilon^{\mu\gamma} T_{\gamma\nu})\xi_{-}^{\nu})\\ \notag
                        + & \difop{24}( \partial_{\mu}(\epsilon^{\mu\gamma} T_{\gamma\nu})\xi_{+}^{\nu})\\ \notag
                          + & \difop{25}( \partial_{\mu}(\epsilon^{\mu\gamma} T_{\gamma\nu})\xi_{-}^{\nu}),
\end{align}
}
with 
\begin{alignat*}{2} 
\cond{1} &=3\, ,\qquad & \conp{1}&=1 \, ,\\
\cond{2} &=3\, ,\qquad &\conp{2}&=5\, , \\
\cond{3} &=3\, ,\qquad &\conp{3}&=3\, , \\
\cond{4} &=3\, ,\qquad &\conp{4}&=1\, , \\
\cond{5} &=3\, ,\qquad &\conp{5}&=3\, , \\
\cond{6} &=3\, ,\qquad &\conp{6}&=1\, , \\
\cond{7} &=3\, ,\qquad &\conp{7}&=-1\, , \\
\cond{8} &=3\, ,\qquad &\conp{8}&=1\, , \\
\cond{9} &=3\, ,\qquad &\conp{9}&=-1\, , \\
\cond{10} &=2\, ,\qquad &\conp{10}&=-3\, , \\
\cond{11} &=2\, ,\qquad &\conp{11}&=4\, , \\
\cond{12} &=2\, ,\qquad &\conp{12}&=2\, , \\
\cond{13} &=2\, ,\qquad &\conp{13}&=0\, , \\
\cond{14} &=2\, ,\qquad &\conp{14}&=2\, , \\
\cond{15} &=2\, ,\qquad &\conp{15}&=0\, , \\
\cond{16} &=2\, ,\qquad &\conp{16}&=-2\, , \\
\cond{17} &=1\, ,\qquad &\conp{17}&=3\, ,\\
\cond{18} &=1\, ,\qquad &\conp{18}&=1\, , \\
\cond{19} &=3\, ,\qquad &\conp{19}&=-1\, , \\
\cond{20} &=3\, ,\qquad &\conp{20}&=3\, , \\
\cond{21} &=3\, ,\qquad &\conp{21}&=1\, , \\
\cond{22} &=3\, ,\qquad &\conp{22}&=1\, , \\
\cond{23} &=3\, ,\qquad &\conp{23}&=-1\, , \\
\cond{24} &=2\, ,\qquad &\conp{24}&=0\, , \\
\cond{25} &=2\, ,\qquad &\conp{25}&=2\, . \\
\end{alignat*}
Note that due to the stress-tensor conservation equation, we have not allowed contraction between $\partial^{\mu}$ and $T^{\mu\nu}$ since it vanishes.
We can once again extract the exact coefficient by a comparison with the three-point function. At large distance and when acting on zero-momentum eigenstates, we find
\bea
E \eps^{\mu\nu} T_{\mu\nu} &=& \frac{\pi q^0}{4} \Bigg( \frac{5}{3}\frac{7\hat{a}+2\hat{b}-\hat{c}}{c_T} \eps^{\mu\nu}T_{\mu\nu} +10 \frac{13\hat{a}+4\hat{b}-3\hat{c}}{c_T} \xi_+^\mu T_{\mu\nu} \epsilon^{\nu\rho}(\xi^-_\rho-\xi^+_\rho) \notag \\
 &-&\frac{15}{6} \frac{81\hat{a}+32\hat{b}-20\hat{c}}{c_T} \xi_+^\mu\xi_+^\nu T_{\mu\nu} \epsilon^{\rho\sigma}(\xi^-_\rho-\xi^+_\rho)(\xi^-_\sigma-\xi^+_\sigma) \Bigg) \,,
\eea
with
\be
c_T= \pi^2\frac{14\hat{a}-2\hat{b}-5\hat{c}}{3} \,.
\ee
This reproduces the expectation of the transfer matrix \rref{transferMatrix}, with the coefficients in agreement with \cite{Hofman:2008ar}.

\section{2d: Two scalars and two stress tensors\label{subsec:TPhi}}
The $n-$point functions of stress-tensors with themselves or with scalar fields can be computed exactly using Ward identites \cite{Belavin:1984vu} that are recalled here for convenience 
\begin{multline}
\langle T(\xi)T(x_1)\dots T(x_M)\phi_1(z_1)\dots\phi_N(z_N)\rangle = \\ \left\lbrace\sum_{i=1}^{N}\left[\frac{\Delta_i}{(\xi-z_i)^2}+\frac{1}{\xi-z_i}\frac{\partial}{\partial z_i}\right]+\sum_{j=1}^{M}\left[\frac{2}{(\xi-x_j)^2}+\frac{1}{\xi-x_j}\frac{\partial}{\partial x_j}\right]\right\rbrace
\langle T(x_1)\dots T(x_M)\phi_1(z_1)\dots\phi_N(z_N)\rangle\\
+\sum_{j=1}^{M}\left[\frac{c/2}{(\xi-x_j)^4}\right]
\langle T(x_1)\dots T(x_{j-1})T(x_{j+1})\dots T(x_M)\phi_1(z_1)\dots\phi_N(z_N)\rangle.
\label{2dWardId}
\end{multline}
We can use the two-point function of two scalars of conformal weight $h$ \be 
\langle \phi(z_1)\phi(z_2)\rangle = \frac{1}{(z_1-z_2)^{2h}},
\ee
as well as the Ward identity \eqref{2dWardId} to compute $\langle T(z_1)\phi(z_2)T(z_3)\phi(z_4)\rangle$, which is given as
\be
\braket{T(z_1)\phi(z_2)T(z_3)\phi(z_4)}=\frac{1}{2}\frac{c}{z_{24}^{2h}z_{13}^4} + \frac{h(hz_{13}^2z_{24}^2-2z_{12}z_{23}z_{14}z_{34})}{z_{24}^{2h-2}z_{12}^2z_{13}^2z_{23}^2z_{14}^2z_{34}^2}\, .\label{TphiTphi}
\ee
It is then straightforward to integrate twice to obtain
\begin{align}
\braket{\mathcal{E}_1\phi(z_2)\mathcal{E}_3\phi(z_4)}= \int\langle T(z_1)\phi(z_2)T(z_3)\phi(z_4)\rangle dz_1dz_3 
%& \int\int (z_{12}z_{34})^{-2-h} \left(\frac{z_{24}}{z_{13}}\right)^{2-h}\left[ \eta ^h \left(h^2+\left(2 h^2-2 h\right) \eta +\dots\right)\right]dz_1dz_3\nonumber\\
%   =& \int\int\left( h^2 \frac{1}{z_{12}^{2}}\frac{1}{z_{34}^{2}}\frac{1}{z_{13}^{2}}z_{24}^{2-2h} +2h(h-1)\frac{1}{z_{12}}\frac{1}{z_{34}}\frac{1}{z_{13}^{3}}z_{24}^{1-2h}+\dots\right)dz_1dz_3\\
%   =& -\frac{6h^2}{(z_2-z_4)^{2+2h}}+\frac{2h(h-1)}{(z_2-z_4)^{2+2h}}\\
    =& -(4\pi^2)\frac{2h(2h+1)}{(z_2-z_4)^{2+2h}}.\label{eq:TpTpFin}
\end{align}

\subsection{Conformal block expansion}
The result \eqref{TphiTphi} can be recast as  
   \begin{align}
\langle T(z_1)\phi(z_2)T(z_3)\phi(z_4)\rangle = 
%&z_{12}^{-2-h} \left(\frac{z_{14}}{z_{13}}\right){}^{2-h}
  % \left(\frac{z_{24}}{z_{14}}\right){}^{2-h} z_{34}^{-2-h}\mathcal{F}_{T\phi T\phi}(\eta),\\
   &z_{12}^{-2-h} \left(\frac{z_{24}}{z_{13}}\right){}^{2-h} z_{34}^{-2-h}\mathcal{F}_{\phi_1\phi_2\phi_3\phi_4}(\eta),
\end{align}
with
%The result for the conformal partial wave is also given in \cite{Osborn} and it reads 
\begin{align}
\mathcal{F}_{T\phi T\phi}(\eta) & = \frac{1}{2}c\eta^{h+2} + \frac{\eta^h}{(1-\eta)^{2}}(h^2-2h\eta(1-\eta))\label{eq:WardScalOsbo}\\
& = \sum_{n=0}^{\infty}C_n \eta^{h+n}F(2h+n-2,n+2;2h+2n;\eta)\, .\label{eq:ConfBlockOsb}
\end{align}
Here, \eqref{eq:WardScalOsbo} is the result one get by direct computation using the Ward identity \eqref{2dWardId}, as we did in \eqref{TphiTphi} while \eqref{eq:ConfBlockOsb} is the conformal block expansion. The coefficients $C_n$ can be found in \cite{Osborn:2012vt}. %with the expansion coefficient given by 
%\begin{align}
%C_n =&\left(\frac{1}{12}(n_1)_3(2h)_{n+1}c + 2h(n(n+2h-1)+1)(2h-2)_n\right)(-1)^n\frac{(2h-2)_n}{(2h-2)_{2n+1}}\nonumber\\
%&+ h((n+1)(2h-2+n)-2)(n+1)!\frac{(2h-2)_n}{(2h-2)_{2n+1}}
%\end{align}
 When expanding $\mathcal{F}_{T\phi T\phi}$ for small $\eta$, we get contributions of the form   
\begin{align}
z_{12}^{-2-h} \left(\frac{z_{24}}{z_{13}}\right)^{2-h} z_{34}^{-2-h}\eta^m 
%= & z_{12}^{-2-h} \left(\frac{z_{24}}{z_{13}}\right)^{2-h} z_{34}^{-2-h}\left(\frac{z_{12}z_{34}}%{z_{13}z_{24}}\right)^m\\
=& \frac{1}{z_{12}^{2+h-m}}\frac{1}{z_{34}^{2+h-m}}z_{24}^{2-h-m}z_{13}^{-2+h-m}\, .\label{eq:Kinematic4Dfactor}
\end{align}
% Let us play the same game as before and calculate the contributions from \eqref{eq:Kinematic4Dfactor}. We got, with $p = h,h+1,\dots$ 
%\begin{align}
%\int\int \frac{1}{z_{12}^{2+h-p}}\frac{1}{z_{34}^{2+h-p}}z_{24}^{2-h-p}z_{13}^{-2-h+p} dz_1dz_3
%\end{align}
When extracting the residues as $z_1\rightarrow z_2$ and $z_3 \rightarrow z_4$, only the terms with $m=h$ and $m=h+1$ will be non-vanishing while all contributions with $m\geq h+2$ vanish.
% The contribution for $p=h$ is given by 
%\begin{align}
%%\int\int \frac{1}{z_{12}^{2+h-h}}\frac{1}{z_{34}^{2+h-h}}z_{24}^{2-h-h}z_{13}^{-2-h+h} dz_1dz_3 &=
% \int\int \frac{1}{z_{12}^{2}}\frac{1}{z_{34}^{2}}\frac{1}{z_{13}^{2}}z_{24}^{2-2h} dz_1dz_3 = -(4\pi^2)\frac{6}{(z_2-z_4)^{2+2h}},
%\end{align} 
%and for $p=h+1$ by 
%\begin{align}
%\int\int \frac{1}{z_{12}}\frac{1}{z_{34}}\frac{1}{z_{13}^{3}}z_{24}^{1-2h} dz_1dz_3 = (4\pi^2)\frac{1}{(z_2-z_4)^{2+2h}},
%\end{align} 

Expanding the exact Ward identity result \eqref{eq:WardScalOsbo} for small $\eta$ yields 
\begin{align}
\mathcal{F}_{T\phi T\phi}(\eta) 
%& = \frac{1}{2}c\eta^{h+2} + \frac{\eta^h}{(1-\eta)^{2}}(h^2-2h\eta(1-\eta))\\
& = \eta ^h \left(h^2+\left(2 h^2-2 h\right) \eta +\left(3 h^2-2 h+\frac{c}{2}\right) \eta ^2+\left(4 h^2-2
   h\right) \eta ^3+\mathcal{O}\left(\eta
   ^4\right)\right),
\end{align}
which once integrated gives
\begin{align}
\braket{\mathcal{E}_1\phi(z_2)\mathcal{E}_3\phi(z_4)}= \int\langle T(z_1)\phi(z_2)T(z_3)\phi(z_4)\rangle dz_1dz_3 
%& \int\int (z_{12}z_{34})^{-2-h} \left(\frac{z_{24}}{z_{13}}\right)^{2-h}\left[ \eta ^h \left(h^2+\left(2 h^2-2 h\right) \eta +\dots\right)\right]dz_1dz_3\nonumber\\
%   =& \int\int\left( h^2 \frac{1}{z_{12}^{2}}\frac{1}{z_{34}^{2}}\frac{1}{z_{13}^{2}}z_{24}^{2-2h} +2h(h-1)\frac{1}{z_{12}}\frac{1}{z_{34}}\frac{1}{z_{13}^{3}}z_{24}^{1-2h}+\dots\right)dz_1dz_3\\
%   =& -\frac{6h^2}{(z_2-z_4)^{2+2h}}+\frac{2h(h-1)}{(z_2-z_4)^{2+2h}}\\
    =& -(4\pi^2)\frac{2h(2h+1)}{(z_2-z_4)^{2+2h}}.\label{eq:TpTpFin}
\end{align}  
This result naturally matches the one obtained by directly integrating the exact result \eqref{eq:WardScalOsbo}. 
\subsection{OPE computation}
We can also reproduce this result using the OPE of the stress-tensor with a scalar field $\phi$, which is 
\be 
T(z)\phi(w) \sim \frac{h\phi(w)}{(z-w)^2} + \frac{\partial_w \phi(w)}{(z-w)}+\dots
\ee
The OPE between $\epsilon(z)$ and $\phi(w)$ will simply project onto $\partial\phi(w)$. We then obtain 
\be 
\braket{\phi(z_2)\mathcal{E}_1\mathcal{E}_3\phi(z_2)} = 4\pi^2 \partial_{z_2}\partial_{z_4}\braket{\phi(z_2)\phi(z_4)} = - (4\pi^2)\frac{2h(2h+1)}{(z_2-z_4)^{2h+2}}\, .
\ee
%Using this result twice in the four-point function yields 
%\begin{align}
%\langle T(z_1)\phi(z_2)T(z_3)\phi(z_4)\rangle 
%%=&\frac{h^2}{z_{12}^2z_{34}^2}\langle \phi(z_2)\phi(z_4)\rangle \\
%%& + \frac{h}{z_{12}^2(z_{34}}\partial_{z_4}\langle \phi(z_2)\phi(z_4)\rangle\\
%%& + \frac{h}{(z_1-z_2)(z_3-z_4)^2}\partial_{z_2}\langle \phi(z_2)\phi(z_4)\rangle\\
%%& + \frac{1}{(z_1-z_2)(z_3-z_4)}\partial_{z_2}\partial_{z_4}\langle \phi(z_2)\phi(z_4)\rangle\\
%&=\frac{h^2}{z_{12}^2z_{34}^2z_{24}^{2h}}  + \frac{2h^2}{z_{12}^2z_{34}z_{24}^{2h+1}}- \frac{2h^2}{z_{12}z_{34}^2z_{24}^{2h+1}} - \frac{2h(2h+1)}{z_{12}z_{34}z_{24}^{2h+2}}.\nonumber\\
%\end{align}
%When performing the first $dz_1$ integral, only the two single order poles remain. Performing the second integral, only the term with a single order pole remains and the final answer is  
%\begin{align}
%\int \langle T(z_1)\phi(z_2)T(z_3)\phi(z_4)\rangle dz_1dz_3
%=& .
%\end{align}
This result is identical to \eqref{eq:TpTpFin}.  
%\begin{align}
%\int \langle T(z_1)\phi(z_2)T(z_3)\phi(z_4)\rangle dz_1
%=&  - (-2\pi i)\left(\frac{2h^2}{(z_3-z_4)^2(z_2-z_4)^{2h+1}} - \frac{2h(2h+1)}{(z_3-z_4)(z_2-z_4)^{2h+2}}\right).
%\end{align}

\bibliographystyle{ytphys}
\bibliography{ref}

\end{document}